\documentclass[useAMS,usenatbib]{mn2e}
\usepackage{bethmacros}
\usepackage{graphicx}
\usepackage{amssymb}

\def \spose #1{\hbox to 0pt{#1\hss}}
\def \lta {\mathrel{\spose{\lower 3pt\hbox{$\mathchar"218$}}
     \raise 2.0pt\hbox{$\mathchar"13C$}}}
\def \gta {\mathrel{\spose{\lower 3pt\hbox{$\mathchar"218$}}
     \raise 2.0pt\hbox{$\mathchar"13E$}}}
\def \kms {{km s$^{-1}$}}

\def \gas {\textsc{gasoline}}
\def \gastwo {\textsc{ESF-gasoline2}}

\def \msun {\ifmmode \rm M_{\odot} \else $\rm M_{\odot}$ \fi}

\title[Constant Disc Gas Mass]{NIHAO III: The constant disc gas mass conspiracy }

\author[Stinson et al.]{G.\,S. Stinson$^{1}$\thanks{Email: stinson `at' mpia.de}, A. A. Dutton$^1$, L. Wang$^{1,2}$, A. V. Macci\`o$^1$, J. Herpich$^1$, 
\newauthor{J. D. Bradford$^3$, T. R. Quinn$^{4}$, J. Wadsley$^{5}$,  B. Keller$^{5}$}
\vspace*{6pt}\\
$^{1}$Max-Planck-Institut f\"ur Astronomie, K\"onigstuhl 17, 69117, Heidelberg, Germany\\
$^{2}$Purple Mountain Observatory, Nanjing, 2 West Beijing Road, Nanjing 210008, China\\
$^3$Astronomy Department, Yale University, New Haven, CT 06520, USA\\
$^{4}$Astronomy Department, University of Washington, Box 351580, Seattle, WA, 98195-1580\\
$^{5}$Department of Physics and Astronomy, McMaster University, Hamilton, Ontario, L8S 4M1, Canada}

\begin{document}
\maketitle
\label{firstpage}

\begin{abstract}
We show that the cool gas masses of galactic discs reach a steady state that lasts many Gyr after their last major merger in cosmological hydrodynamic simulations.  The mass of disc gas, M$_{\rm gas}$, depends upon a galaxy halo's spin and virial mass, but not upon stellar feedback.  Halos with low spin have high star formation efficiency and lower disc gas mass.  Similarly, lower stellar feedback leads to more star formation so the gas mass ends up nearly the same irregardless of stellar feedback strength. Even considering spin, the M$_{\rm gas}$ relation with halo mass, M$_{200}$ only shows a factor of 3 scatter.  The M$_{\rm gas}$--M$_{200}$ relation show a break at M$_{200}=2\times10^{11} \msun$ that corresponds to an observed break in the M$_{\rm gas}$--M$_\star$ relation.
The constant disc mass stems from a shared halo gas density profile in all the simulated galaxies.  In their outer regions, the profiles are isothermal.  Where the profile rises above $n=10^{-3}$ cm$^{-3}$, the gas readily cools and the profile steepens.  Inside the disc, rotation supports gas with a flatter density profile and at low masses, supernova explosions disrupt the disc.  Energy injection from stellar feedback also provides pressure support to the halo gas to prevent runaway cooling flows.  The resulting constant gas mass makes simpler models for galaxy formation possible, either using a ``bathtub'' model for star formation rates or when modeling chemical evolution.
\end{abstract}

\begin{keywords}
galaxies: formation -- galaxies:ISM -- hydrodynamics -- methods: N-body simulation
\end{keywords}

\section{Introduction}
How gas evolves, especially gas at temperatures well below a galaxy halo's virial temperature, remains one of the central questions of galaxy formation.  The accretion and cooling of gas onto discs makes star formation possible.  There are large reservoirs of hot gas surrounding galaxies \citep{Tumlinson2011,Werk2014} and centrifugally supported discs of cool gas that form stars.  

There are a number of physical processes that can move gas between the disc and halo.  Processes that add gas to discs include: \begin{itemize}
\item \emph{Cold flow accretion} of pristine cold gas through cold, filamentary accretion \citep{Keres2005}
\item \emph{Hot mode accretion}, which happens when gas radiatively cools out of the hot halo onto the disc \citep{Rees1977,White1978,Putman2012,Werk2014} 
\item \emph{Evolved stars}:  Stellar populations lose $\sim40$\% of their initial mass through stellar winds and planetary nebulae \citep{Kalirai2008,Leitner2011}
\item \emph{Stimulated accretion}:  In a process not resolved in current cosmological simulations, metal enriched outflows from the disc can seed the halo with enough metals and turbulence to encourage accretion onto the disc \citep{Fraternali2008,Marinacci2010}
\item \emph{Gas recycling}:  In a similar process to stimulated accretion, simulations show that metal rich gas ejected through stellar winds often falls back onto the disc \citep{Oppenheimer2010,Brook2014,Ubler2014}
\end{itemize}
Sinks that can remove gas from discs include:
\begin{itemize}
\item \emph{Star formation}
\item \emph{Galactic outflows}:  stellar or AGN driven
\item \emph{Photoionization or cosmic rays}:  strong photo ionizing sources photo-evaporate cool gas \citep{Kannan2014}
\end{itemize}

With such a wide variety of processes that add and remove cold gas from galactic discs, it seems like the disc gas mass would not stay constant.  However, there have been a number of recent steady state models of galaxy formation that suppose exactly that \citep{Bouche2010,Bauermeister2010,Dave2012,Lilly2013,Dekel2014}.  Such models present a relatively simple picture of galaxy formation and with such a simple description, they do a remarkably good job reproducing key observables like the evolution of the specific star formation rate.  It is thus worthwhile exploring in numerical simulations: \emph{Does the assumption of constant gas mass hold?}

The dynamical time for the inner regions of galaxies immediately surrounding the disc is only a few percent of the Hubble time.  So, in the absence of mergers, it seems plausible that all the processes listed above might balance to create a quasi-steady state.

There are a variety of observational surveys of the cold gas content in galaxies that provide a basis for comparison with models.  For neutral hydrogen ({\sc Hi}), there have been the HI Parkes All-Sky Survey {\sc hipass} \citep{Meyer2004}, the Arecibo Legacy Fast ALFA ({\sc alfalfa}) Survey \citep{Giovanelli2005}, and the Galex Arecibo {\sc sdss} Survey ({\sc gass})  \citep{Catinella2012}.  The {\sc cold gass} Survey augmented {\sc gass} with the inclusion of CO observations that give some idea about the mass of hydrogen in the molecular phase.  These surveys of the local Universe provide a good census of the cold gas in galactic discs.  

There have been observations of CO throughout cosmic history back to $z\sim3$ \citep{Daddi2010,Genzel2010,Tacconi2013}.  However, the 21 cm emission from {\sc Hi} is too faint to see beyond the local Universe.  For this reason, more powerful instruments like the Square Kilometer Array (SKA) \citep{SKA} are required to nail down the total gas mass evolution of galaxies.  Using the spectra of absorption lines from distant quasars.  \citet{Prochaska2005} found that the cosmic density of {\sc Hi}, $\Omega_{\rm HI}$, is constant \citep[also see][]{Crighton2015}.

Relatively simple models have been able to match the data well.  Semi-empirical methods estimate the gas content of galaxies based on derived star formation rates from abundance matching \citep{Lu2014,Popping2015}.  Such a model extrapolates H$_2$ masses from the star formation rate based on the Kennicutt-Schmidt star formation--gas surface density relationship \citep{Schmidt1959,Kennicutt1998,Bigiel2008}.  The model further extrapolates {\sc Hi} masses from gas pressure arguments \citep{Blitz2006}.  The model does a good job matching H$_2$ masses at redshifts $z\sim2$.  These analytic models imply little evolution of the relationship between M$_{\rm HI}$ and M$_{\rm vir}$ across cosmic time, though the exact location of that gas unknown.  Some may be in discs, but much could be in the cosmic web or surrounding galaxies.

Similar estimates have been applied to other galaxy formation models to show the gas content and evolution of a wide range of galaxy models.  Using a model for galaxy formation based on angular momentum conservation, \citet{Dutton2010} found constant cold gas mass with a gradual shift from molecular to atomic hydrogen as galaxies grow larger and their surface densities decrease.  \citet{Lagos2011} showed how gas evolves using the GALFORM semi-analytic model from Durham.  \citet{Popping2014} examined gas properties of the \citet{Somerville1999} semi-analytic model and found a similar result.  \citet{Popping2015} and \citet{Lu2014} used a semi-empirical approach to constrain the gas evolution in galaxies much like has been done with stars using abundance matching and halo occupation distributions.  In their semi-empirical work, \citet{Popping2015} points out that the {\sc Hi} content of galaxies does not evolve much, exactly like we emphasize here.

Numerical simulations of galaxy formation have entered an era in which it is possible to create large samples of realistic galaxies \citep{Aumer2013,Hopkins2014,Vogelsberger2014,Schaye2015, Wang2015}.  \citet{Genel2014} uses a simple formulation to estimate the H$_2$ fraction in the Illustris galaxies and finds agreement with observations at $z=0$, but not at $z=2$.  Some studies have focused on the gas content and distribution in galaxy simulations.  In a paper on the stellar mass metallicity relation, \citet{Finlator2008} found minimal changes in disc gas mass and \citet{Dekel2013} found the same in a paper focused on comparing simulations with analytic models.  \citet{Nelson2015} examines several zoom regions at $z\sim2$ and finds that gas distributions are asymetric, particularly as resolution increases. 

We examine the cold gas properties and evolution in Numerical Investigation of a Hundred Astrophysical Objects, NIHAO \citep{Wang2015}.  The NIHAO project strives to simulate 100 galaxies across a wide mass range, $10^5<$M$_\star$/M$_\odot<10^{11}$, using hydrodynamics in a cosmological context.  The NIHAO simulations are based on the star formation and feedback that was developed for the Making Galaxies in a Cosmological Context project \citep[MaGICC,][]{Stinson2013}.  The MaGICC simulations used physically motivated feedback to create a realistic galaxy of about Milky Way mass.  The NIHAO project extends the range of simulations down to the smallest mass galaxy that will form stars, $\sim10^5 \msun$.  Using the same stellar physics across the entire mass range, \citet{Wang2015} find that the stellar mass of all galaxies agree with results from abundance matching.  This makes the NIHAO sample ideal for exploring the physics of galaxy formation.

Here, we examine the gas, especially the cold gas, in the ensemble of the NIHAO galaxies.  We find that despite all the complicated physics, the cool disc gas mass remains largely constant within a mass range of $10^9<$M$_\star$/M$_\odot<10^{10.5}$ and at moderate spin parameters, $0.03<\lambda<0.06$.  We explore what the ramifications of the constancy might be for studying galaxy evolution.  In \S \ref{sec:sims}, we provide an overview of the simulations used to find this result.  \S \ref{sec:results} shows the constant gas mass and how it relates to halo virial mass.  \S \ref{sec:uni} discusses the gas density profiles that form in the simulations.  \S \ref{sec:discussion} provides a context for this finding in recent theoretical models, observations of gas in galaxies, and some possible implications for future observations.

\section{Simulations}
\label{sec:sims}
The simulations analyzed in this paper are taken from the NIHAO project \citep{Wang2015}.  The NIHAO galaxies come from (20 $h^{-1}$ Mpc)$^3$ and (60 $h^{-1}$ Mpc)$^3$ cubes from \citet{Dutton2014}.  They use parameters from \citet{Planck2014}: $\Omega_{\rm m}$=0.3175, $\Omega_\Lambda$=0.6825, $\Omega_{\rm bary}$=0.049, $H_0=67.1$ \kms Mpc$^{-1}$, $\sigma_8=0.8344$.   

\subsection{ESF-Gasoline2}
In their paper describing superbubble feedback, \citet{Keller2014} described the updates they made to the N-body SPH solver \textsc{gasoline} \citep{Wadsley2004}.  While we do not use their superbubble feedback, we employ their modified version of hydrodynamics that removes spurious numerical surface tension and improves multiphase mixing.  We thus refer to our version of \textsc{gasoline} as \gastwo, where {\sc esf} stands for the ``early stellar feedback'' that is described below.  The biggest differences to the hydrodynamics in \gastwo\, come from the small change \citet{Ritchie2001} proposed for calculating $\frac{P}{\rho^2}$.  \citet{Ritchie2001} also proposed modifying the density calculation to use equal pressures, but we do \emph{not} use those densities in the simulations described here.

Diffusion of quantities like metals and thermal energy between particles has been implemented as described in \citet{Wadsley2008}.  Metal diffusion is used, but thermal diffusion is not used because it is incompatible with the blastwave feedback that delays cooling.  \gastwo\, includes several other changes to the hydrodynamic calculation.  The \citet{Saitoh2009} timestep limiter was implemented so that cool particles behave correctly when a hot blastwave hits them.  To avoid pair instabilities, \gastwo\, uses the Wendland $C2$ function for its smoothing kernel \citep{Dehnen2012}.  The treatment of artificial viscosity has been modified to use the signal velocity as in \citet{Price2008}.   

The cooling is as described in \citet{Shen2010} and was calculated using \textsc{cloudy} (version 07.02; \citet{Ferland1998}) including photoionization and heating from the \citet{Haardt2005} UV background, Compton cooling, and hydrogen, helium and metal cooling from 10 to $10^9$ K.  In the dense, interstellar medium gas, we do not impose any shielding from the extragalactic UV field as the extragalactic field is a reasonable approximation in the interstellar medium.  

\subsection{Star Formation and Feedback}
\label{sec:feedback}
The simulations use a common recipe for star formation described in \citet{Stinson2006} that we summarize here.  Stars form from cool ($T < 15,000$ K), dense gas.  The metal cooling readily produces dense gas, so the star formation density threshold is set to the maximum density at which gravitational instabilities can be resolved, $\frac{32 M_{gas}}{\epsilon^3}$($n_{th} > 9.3$ cm$^{-3}$), where $M_{gas}=2.2\times10^5$ M$_\odot$ and $\epsilon$ is the gravitational softening (310 pc).  Such gas is converted to stars according to the equation
\begin{equation}
\frac{\Delta M_\star}{\Delta t} = c_\star \frac{M_{gas}}{t_{dyn}} .
\end{equation}
Here, $\Delta M_\star$ is the mass of the star particle formed, $\Delta t$ is the timestep between star formation events, $8\times10^5$ yr in these simulations, and $t_{dyn}$ is the gas particle's dynamical time.  $c_\star$ is the efficiency of star formation, i.e. the fraction of gas that will be converted into stars during $t_{dyn}$. 

The simulations use the same feedback described in \citet{Stinson2013}.  In this scheme, the stars feed energy back in two epochs.  The first epoch, ``pre-SN feedback'' (ESF), happens before any supernovae explode.  It represents stellar winds and photoionization from the bright young stars.  
In \gas\,, as in \citet{Stinson2013}, the pre-SN feedback consists of 10\% of the total stellar flux being ejected from stars into surrounding gas ($2\times10^{50}$ erg of thermal energy per M$_\odot$ of the entire stellar population).   We show \gastwo\, simulations that use $\epsilon_{ESF}$=10\%.  Because of the increased mixing in \gastwo, the simulations required more stellar feedback to have their star formation limited to the abundance matching value.  Thus, we also show results from using $\epsilon_{ESF}$=13\%, which gives a better match to \citet{Behroozi2013} abundance matching results.  Radiative cooling is left \emph{on} for the pre-SN feedback.
 
The second epoch starts 4 Myr after the star forms, when the first supernovae start exploding.  Only supernova energy is considered as feedback in this second epoch.  
Stars $8$ M$_\odot <$ M$_\star < 40$ M$_\odot$ eject both energy and metals into the interstellar medium gas surrounding the region where they formed.  Supernova feedback is implemented using the blastwave formalism described in \citet{Stinson2006}.  Since the gas receiving the energy is dense, it would quickly be radiated away due to its efficient cooling.  For this reason, cooling is delayed for particles inside the blast region for $\sim30$ Myr.

\subsection{Cool Gas versus HI+H$_2$}
We use two different measures of the cool gas content in the paper, one for comparison with observations (M$_{HI+H_2}$), and one that is more physically relevant, M$_{gas}$(T$<$20000 K).  

For comparison with observations (Fig. \ref{fig:mhimstar}), we show the non-equilibrium calculated \textsc{Hi}.
The non-equilibrium cooling calculation in the simulations tracks the evolution of the atomic ion species of hydrogen and helium \citep{Shen2010}.  We report the neutral hydrogen mass as {\sc Hi}+H$_2$.  The simulations do not resolve densities high enough for H$_2$ to form, so we are not able to subtract the H$_2$ mass from our neutral gas masses.  \citet{Saintonge2011} shows that M$_{\rm H_2}$ are at most equal to M$_{\rm HI}$, so in some cases, it is possible that our masses should be divided by a factor of 2.  

{\sc Hi} observations only represent  emission from the 21 cm transition that only exists at column densities above $10^{18}$ cm$^{-2}$.  It is possible that there is a small contribution to the simulated M$_{HI}$ from lower column density gas, but the UV background should ionize that gas.

As a measure of the total virialized halo, we use everything contained within the surface where the mean density is 200$\rho_c$.  We refer to this radius as $r_{200}$ and the mass inside it as M$_{200}$.  We find that M(T$<20000$ K) is more tightly correlated with M$_{\rm 200}$ than M$_{\rm HI+H_2}$, so for all plots other than Fig. \ref{fig:mhimstar}, we report our findings in terms of M(T$<$20000 K).

\section{Results}
\label{sec:results}
Our study of the gas mass in galaxies starts with a comparison with observations to check whether a detailed analysis is worthwhile.  Since the correspondence with observations is good, we investigate how the gas mass evolves and examine the relationship with halo mass.  We also try to inform the trend we sees with some tests of galaxies as a function of spin parameter and using various recipes for feedback.
\subsection{Comparison with Observations}
\begin{figure}
\resizebox{9cm}{!}{\includegraphics{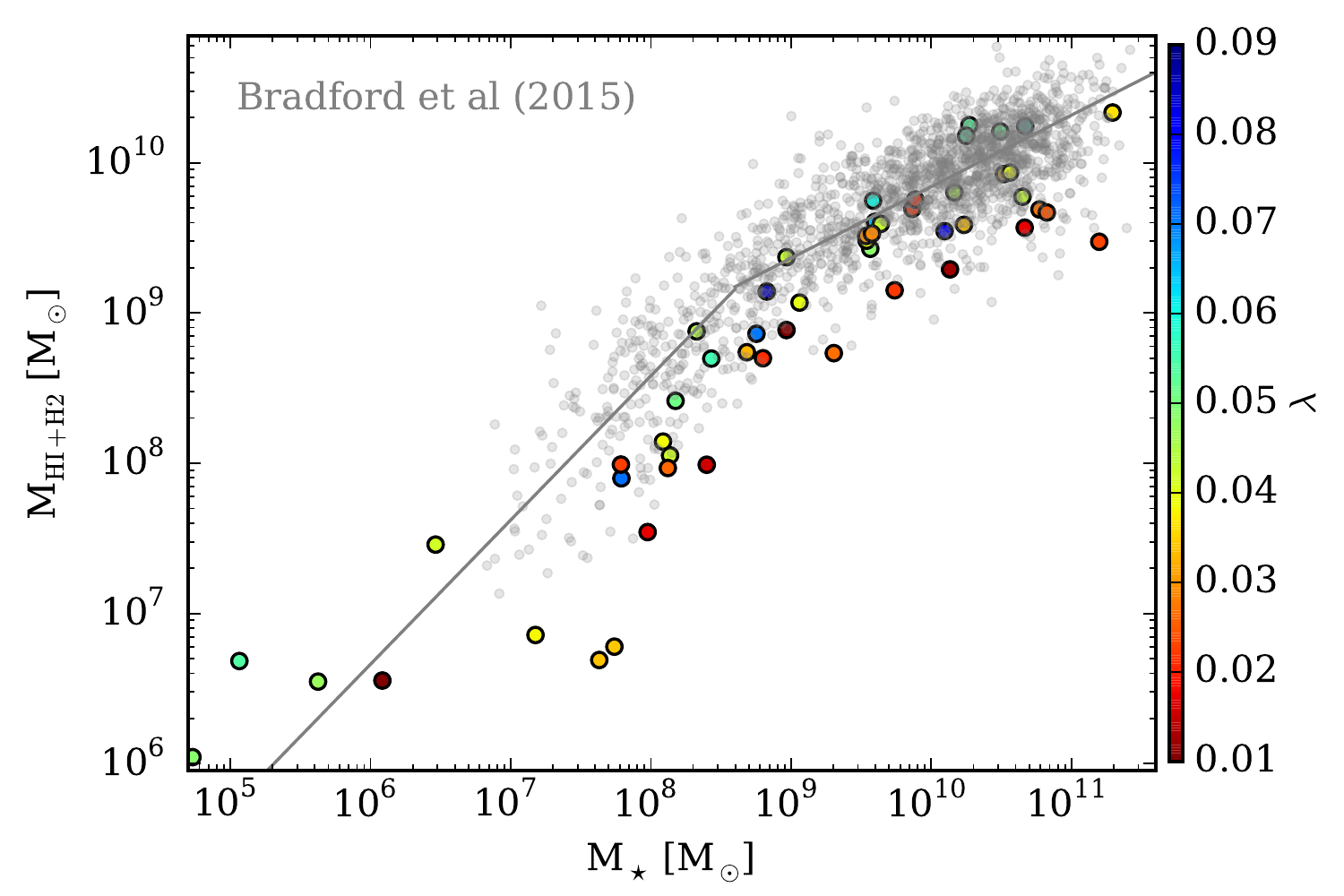}}
 \caption[Comparison with Observations]{ Simulated {\sc Hi} masses as a function of stellar mass compared with an observed sample of galaxies from \citet{Bradford2015}.  The simulated points are coloured according to the total spin parameter at $z=0$ of the entire galactic halo.  Low spin (red) galaxies have lower gas to star ratios than higher spin galaxies because they are more efficient at turning gas into stars}
\label{fig:mhimstar} 
\end{figure}

Fig. \ref{fig:mhimstar} shows how the present-epoch masses of cold (\textsc{Hi}+H$_2$) gas discs in the simulations compare with observations. The simulation data are shown as the points coloured according to spin.  The grey points are observations from the \citet{Bradford2015} sample of isolated galaxies.  The grey line shows the fit \citet{Bradford2015} make to their observations.  The fit is a double power law with a break at M$_\star=10^{8.6} \msun$.  Below the break, \citet{Bradford2015} find M$_{\rm HI + H_2} \propto$ M$_\star^{1.052}$; above the break, they find M$_{\rm HI + H_2} \propto$ M$_\star^{0.461}$.

The \citet{Bradford2015} {\sc Hi} masses come from 21-cm emission lines of their own low mass galaxy sample and their re-measurements of the {\sc alfalfa} survey 40\% data release \citep{Haynes2011}.  These {\sc Hi} observations have been cross matched to a re-reduction of the Sloan Digital Sky Survey (SDSS DR8), where stellar masses have been measured with \textit{kcorrect} using SDSS \textit{ugriz} photometry, GALEX UV photometry and a \citet{Chabrier2003} IMF.  We have also included their M$_{\rm H_2}$ mass estimates, which are derived from H$\alpha$ star formation rates.  See \citet{Bradford2015} for a complete description of these three mass estimates and the isolation criteria they applied to this sample.

The cold gas masses from the simulation are derived using the non-equilibrium cooling calculation performed in the simulations.  The correspondence between the observations and simulations is notable, especially the break in the relationship.  Both the simulated and observed relation steepen near M$_\star<10^9$ M$_\odot$ \citep[a break first noted in][]{Huang2012}.  In general, the scatter of the observations is larger than the simulations.  

There are observed galaxies with higher {\sc Hi} masses than the simulations at all stellar masses.  Both the \citet{Bradford2015} sample and the {\sc alfalfa} survey are flux limited and therefore biased in detecting the most gas-rich galaxies in the local universe.  At low M$_\star$, the simulations exhibit systematically lower M$_{\rm HI + H_2}$ than the observations.  The observations become less complete at these lower masses.

Another notable trend in Fig. \ref{fig:mhimstar} is that the low spin (red points) galaxies tend more towards the lower-right of the observations.  The lower right on this plot represents higher stellar mass and lower gas mass, thus the gas to star ratio is lower for these galaxies.  In other words, gas is more effectively converted to stars if the galaxy has a low spin parameter \citep[also see][]{Huang2012}.  \S \ref{sec:spin} expands on this finding.

\citet{Catinella2012} point out that around 30\% of galaxies at all stellar masses have no detectable 21 cm emission from {\sc Hi} in their {\sc Hi} blind GASS Survey.  Many of these ``gas free'' galaxies\footnote{The quotes around ``gas free'' refer to the fact that while there is no observable dense cool gas, but there is plentiful lower density hot and cold gas in the galaxy halos surrounding the stars \citep[i.e.][]{Werk2014}.} are in groups or clusters where their {\sc Hi} has been stripped.  Our simulated selection is entirely composed of isolated galaxies that will not experience these environmental effects.  Thus, every high mass galaxy in the simulated sample contains observable gas, similar to what \citet{Bradford2015} observe in their isolated sample.  The selection is necessary for simulations since simulating the hydrodynamics of galaxies flying through hot cluster gas is too computationally demanding at present to maintain the resolution of the individual galaxies.

Fig. \ref{fig:mhimstar} shows that the correspondence between simulations and observations is encouraging enough to lend credence that the simulations can be further analysed to see how gas evolves in real galaxies.  

\subsection{Disc Gas Mass Evolution}
\begin{figure*}
\resizebox{18cm}{!}{\includegraphics{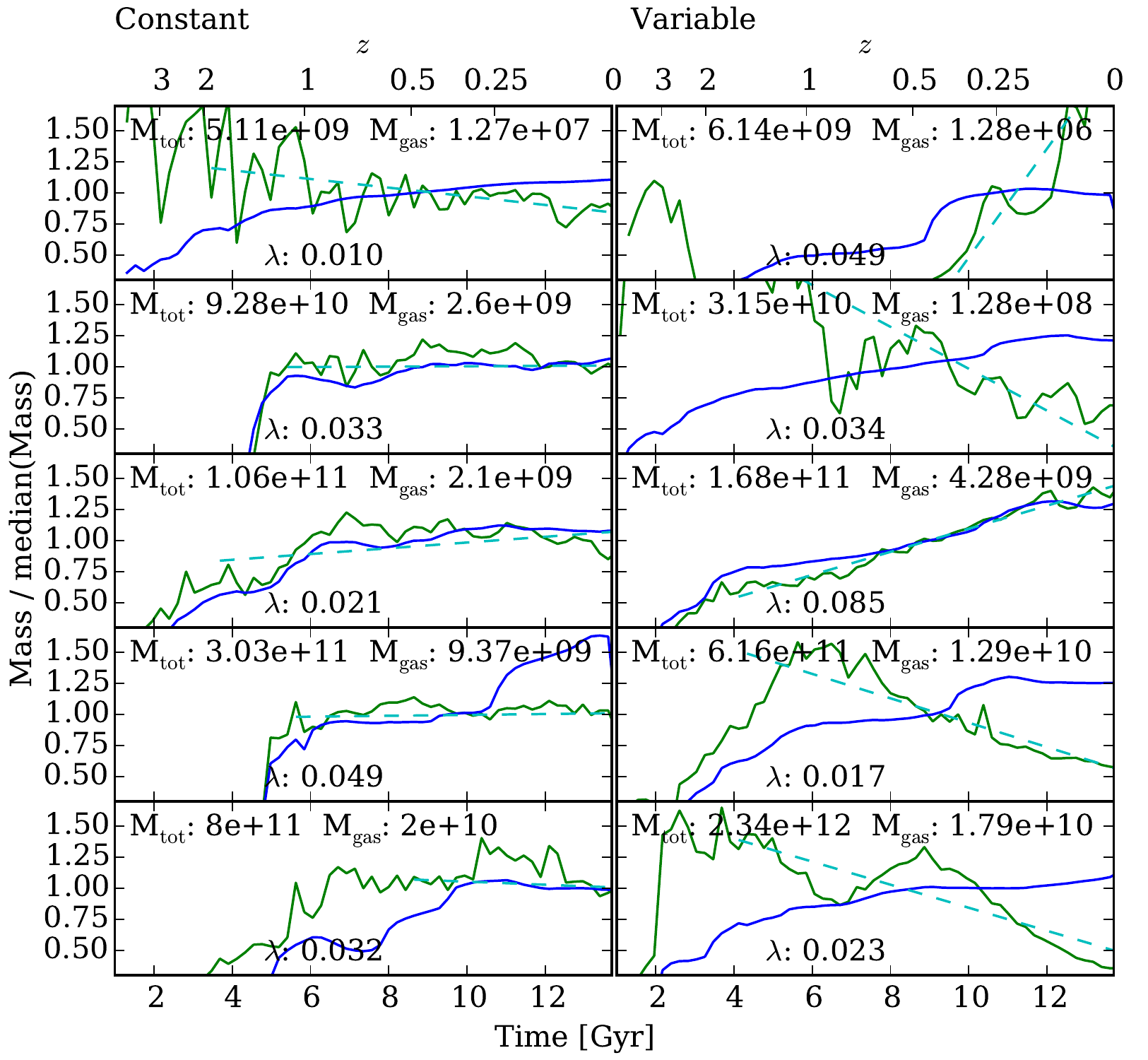}}
 \caption[Disc Gas Mass Evolution]{ Disc gas mass evolution for ten high resolution galaxies from project NIHAO (\emph{green line}).  The values have been normalized by their median value.  The dashed \emph{cyan} line is a linear fit to the evolution of the gas mass from the time of its last major merger until $z=0$.  For comparison, the evolution of the total halo mass is shown as the solid \emph{blue} line.  The galaxies on the left are a representative sample across the mass range of galaxies that maintain their constant disc mass, while the right column shows galaxies in which the disc gas mass gradually evolves. }
\label{fig:ten} 
\end{figure*}

Fig. \ref{fig:ten} shows the evolution of the cool gas (T$<20000$ K, \emph{green line}, \emph{not} the same as M$_{HI}$ shown in Fig. \ref{fig:mhimstar}) mass inside $r_{gal} \equiv 0.2 r_{\rm 200}$ for a sample of ten NIHAO galaxies.  The left column shows galaxies for which the disc gas mass remains within 20\% of the median since the end of the era of major mergers at $z\sim1.5$.  The right column shows galaxies in which the gas mass evolves gradually, though one should notice that the evolution is less than 50\% from the median.  All of the gas mass evolution plots have been normalized so that their median value over this period of evolution is 1.  The range of the y-axis only spans a factor of 4 and all of the galaxies that evolve stay within that range.  The sample of galaxies that maintain a constant gas mass are significantly constant.  Their variation from their median values is never more than 20\%.  

To see how the variation in gas mass compares with the variation of the total halo mass, the total halo mass is shown as the blue line.  The total halo mass has also been normalized so that its median is 1.  The total mass is generally 2 orders of magnitude higher than the gas mass.  The evolution of the total gas mass shows when mergers occur.  The bottom two constant cases show non-negligible merger histories.  The cool gas mass evolves less than the total mass in both of these cases.

The gas and virial masses have been included along with the spin parameter, $\lambda$, at $z=0$ for each of the galaxies.  The spin parameter, $\lambda$, shown is defined using $\lambda'=J/\sqrt{2G M_{200}^3 r_{200}}$ from \citet{Bullock2001} for all the particles, both dark and baryonic, inside $r_{200}$.  Both selections contain galaxies from nearly the entire mass range, though the evolving selection (right panels) includes higher mass galaxies.  It is worth exploring what sort of galaxies maintain constant gas masses and which ones evolve.  

To quantify the evolution of gas masses, linear fits to the gas mass evolution are made from their last major merger until $z=0$.  A major merger is defined as the dark matter mass changing by more than 30\% over 3 snapshots.  The fits are shown in Fig. \ref{fig:ten} as dashed cyan lines.

\begin{figure*}
\resizebox{18cm}{!}{\includegraphics{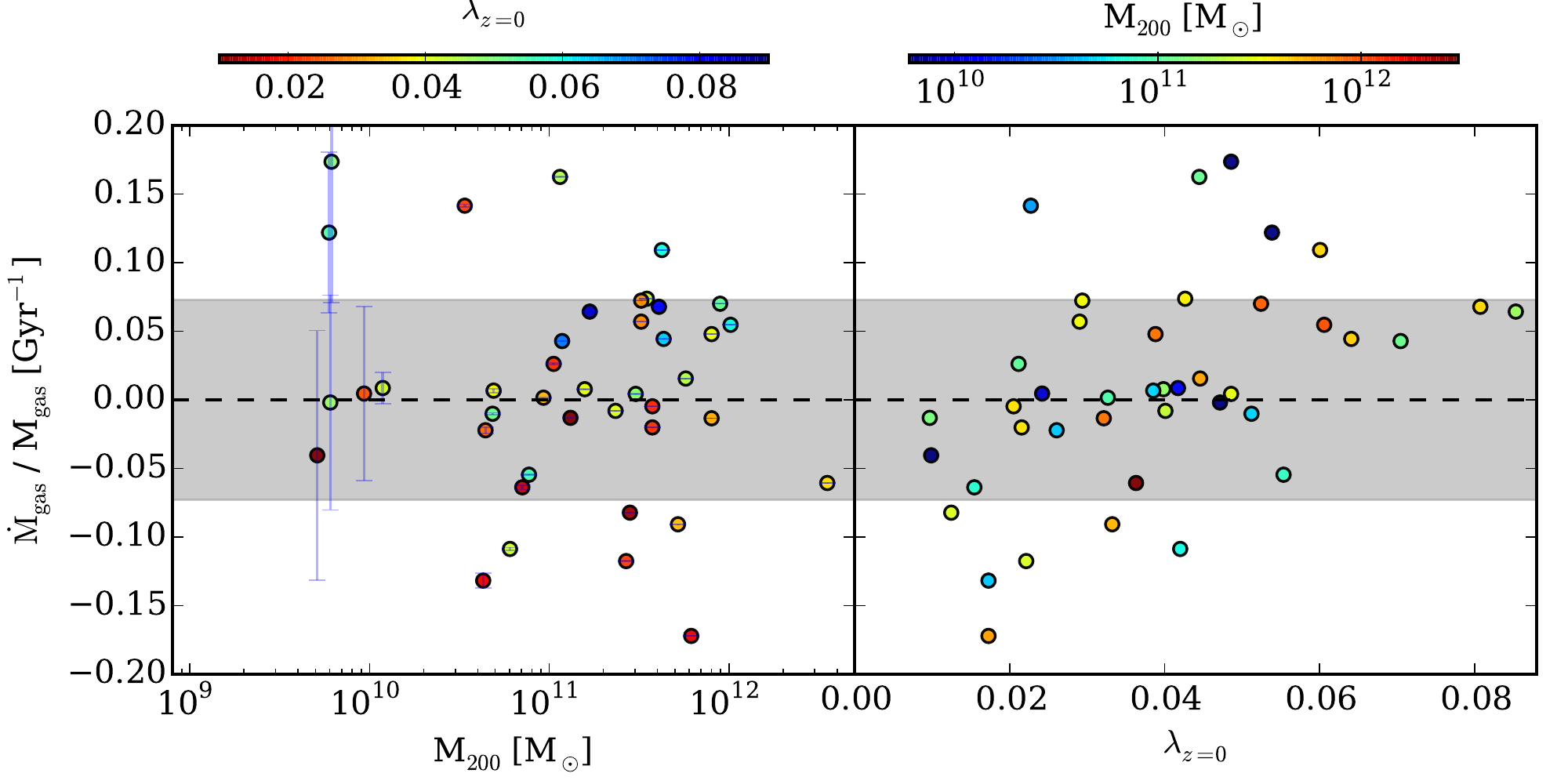}}
 \caption[Gas evolution]{ Disc gas mass evolution from every high resolution galaxies in project NIHAO as a function of halo mass (left panel) and $z=0$ spin parameter (right panel).  $\dot{\rm M}_{gas}$ represents a linear fit to the evolution of cool disc gas mass between $z=1$ and $z=0$.  Those slopes are normalized by the median gas mass during this period to put galaxies of varying masses on even footing.     }
\label{fig:slopes} 
\end{figure*}

The slopes of the linear fits for the entire NIHAO sample are shown in Fig. \ref{fig:slopes}.  The slopes are plotted as a function of halo mass (left) and spin parameter (right).  In the halo mass panel, the point colour represents the halo's spin parameter at $z=0$.  In the spin parameter panel, the point colour represents the halo mass.  

The grey region shows the region in which gas mass changes by less than order unity over the Hubble time.  Most points lie comfortably with this region, emphasizing the point that gas masses remain constant for significant time periods.

Both panels have scatter. The points represent galaxies with a wide variety of merger histories.  As we will see, gas mass typically follows total halo mass, so mergers tend to increase the disc gas mass of galaxies.  Late mergers can also change the spin parameter at the last moment, so $\lambda$ measured at $z=0$ may not be the best time to measure the halo spin, but it is the one that could most readily be observed.  

Despite the complex formation histories for the galaxies, it is still possible to pick out a trend with halo spin: the only galaxies that lose disc mass have $\lambda_{z=0}<0.042$.  That means galaxies that spin less lose cool gas content faster than those that are spinning quickly.  One can understand such a reduction in cool gas with a simple model.  Cool gas is rotationally supported.  In galaxies with less spin, there is less rotational support, so the cool gas collapses towards the centre of the galaxy disc.  As it collapses, it reaches higher surface densities.  Gas turns into stars more readily as it reaches higher surface density \citep{Schmidt1959,Kennicutt1998,Bigiel2008}.  Thus, the cold gas mass starts to decrease.
It is notable that the cosmic median spin parameter is $\sim0.04$ \citep{Bullock2001, Maccio2008}, the spin parameter below which galaxies decrease in cold gas mass.  

There is no trend apparent in cool gas mass as a function of halo mass (left panel).  At all halo masses, there is roughly an equal number of galaxies that lose cool gas mass to those that gain it.  We might expect a decrease in cool gas masses at M$_{200}>10^{12} \msun$ because the virial temperature rises high enough that it takes longer for gas to cool.  Such a decrease would suggest how star formation can be quenched.  Our sample of galaxies does not probe high enough masses to make such a statement.  In the sample of galaxies at $10^{12} \msun$ there is again an even distribution of galaxies gaining and losing mass.  One galaxy at $4\times10^{12} \msun$ has decreasing gas mass but does not make a trend.

\subsection{The Effect of Spin Parameter}
\label{sec:spin}
\begin{figure}
\resizebox{9cm}{!}{\includegraphics{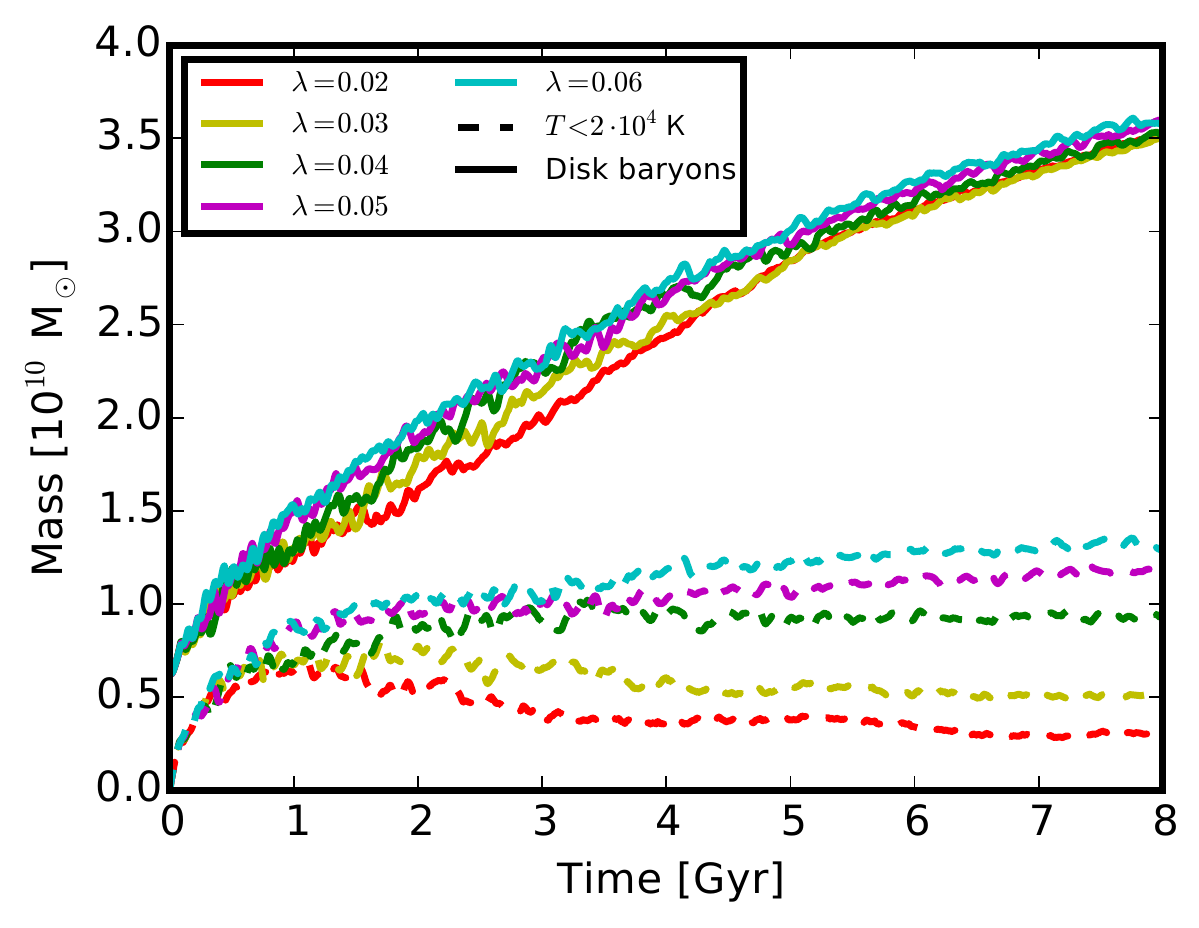}}
 \caption[Disc gas as a function of spin]{ The solid lines show the evolution of the baryonic disc mass in isolated (non-cosmological) simulations of $10^{12}$ M$_\odot$ halos that include spinning gas spheres that cool and collapse to form stars.  The dashed lines show the total baryon content in a disc region with a radius of 20 kpc and 6 kpc thickness.  Because all the halos have the same mass, it is not surprising to see that the same amount of material cools into the disc region.  There is an apparent difference in how much of that material remains as gas.  The gas that is given the most initial spin ($\lambda=0.06$) supports the most gas and the halo with the least spin ($\lambda=0.02$) supports the least gas.  The high spin galaxies also increase in gas mass while the low spin galaxies decrease in cool gas mass.  }
\label{fig:disc_mass_ev} 
\end{figure}

Following up on the spin trend from Fig. \ref{fig:slopes}, we make a closer examination of how spin can affect disc gas mass.  We turn to the simulations from \citet{Herpich2015} that are independent of galaxy merger history.  In their study of isolated collapsing spheres of gas, \citet{Herpich2015} show the halo spin parameter sets the surface density profile for galactic discs.  Fig. \ref{fig:disc_mass_ev} shows another result of \citet{Herpich2015}, namely, the gas fraction of the galaxies depends on how fast the initial sphere is spinning.  The dot-dash lines show the evolution of the total baryonic (gas and stars) mass while the solid lines show the evolution of the gas mass alone.  The range of spin parameters shown ($0.02<\lambda<0.06$) spans the middle of the range of spin parameters found in simulations of cosmological volumes \citep{Bullock2001}.  Even in this small range, one can see the effect spin has on disc gas evolution:  First, the variation in spin parameters leads to a factor of 3 spread in final gas mass.  Second, the high spin galaxies continue to add gas throughout their evolution while the disc gas mass decreases in the low spin galaxies.

Such results are consistent with the analysis in \citet{Mo1998} that shows how the scale radius of discs depends on the spin parameter and virial radius,
\begin{equation}
R_d \propto \lambda r_{200}.
\end{equation}
The evolution of the galaxy sizes is the result of the accretion of higher angular momentum material.  Pristine gas that gets accreted onto the disc at late times has taken the longest time to get there because it started the furthest away.  Thus, it has more angular momentum than the gas that was accreted earlier.  \citeauthor{Brook2012} (2012a) and \citet{Ubler2014} point out that gas ejected from galaxies can also be re-accreted.  Their studies show that ejected gas consists initially of the lowest angular momentum material and when it is re-accreted, it has more angular momentum.

\subsection{Full Population Statistics}

\begin{figure}
\resizebox{9cm}{!}{\includegraphics{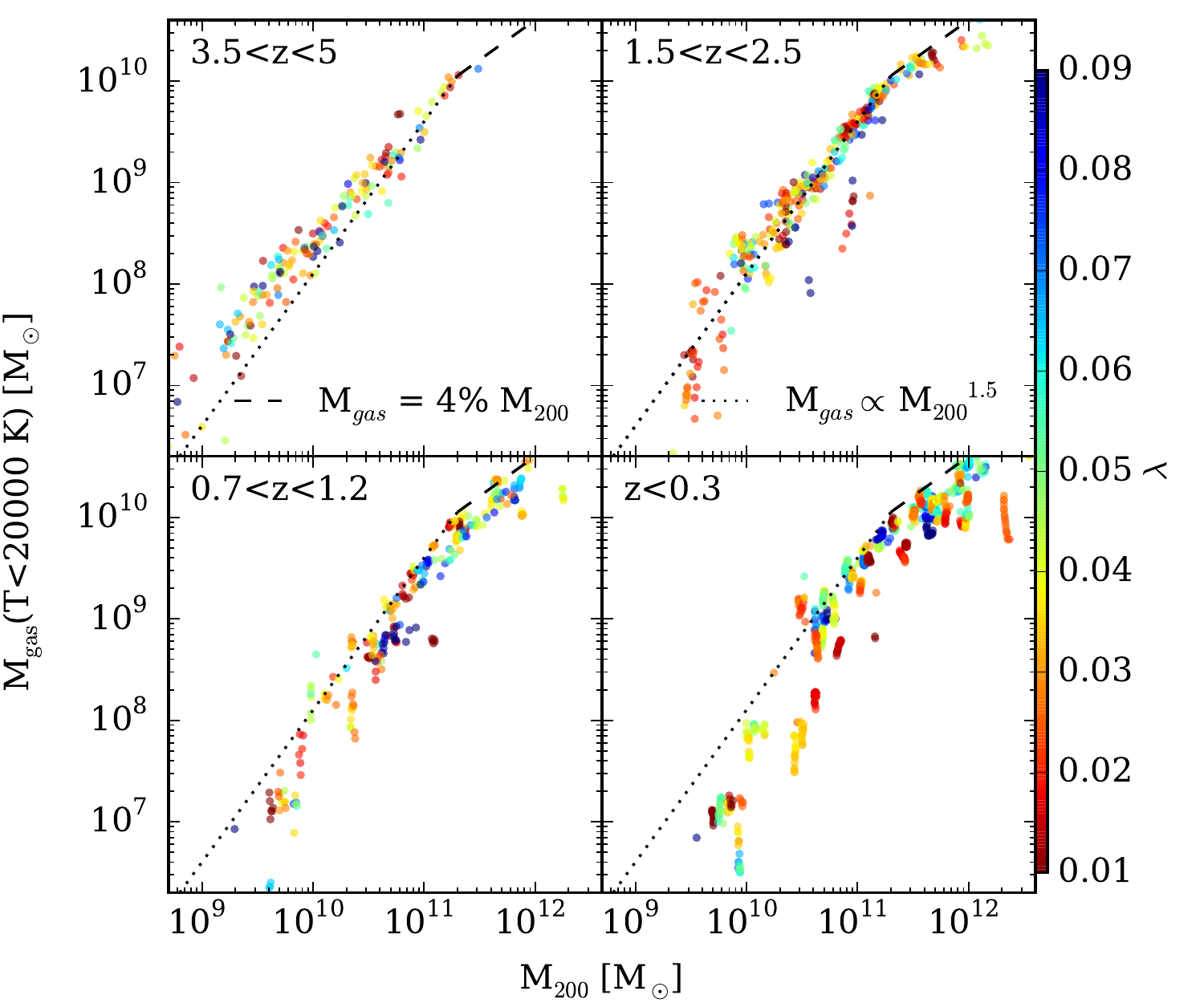}}
 \caption[]{ The cool disc gas mass as a function of M$_{200}$ at 4 different epochs, $z\sim4$, $z\sim2$, $z\sim1$, and $z\sim0$.  The same two lines are drawn on each panel.  At low masses, the dotted line represents the relationship $M_{gas}\propto M_{200}^{1.5}$, while at $M_{200}>2\times10^{11} \msun$, $M_{gas}=0.04 M_{200}$.  The points are coloured according to the spin parameter of the galaxy at the moment the mass is measured.}
\label{fig:lmgas} 
\end{figure}
We have seen disc gas masses remain mostly constant over the course of many Gyr of evolution.  There are some fluctuations and systematic trends depending on the spin parameter but not halo mass.  To gain a more complete understanding of what sets the cool disc gas mass, Fig. \ref{fig:lmgas} shows the cool disc mass as a function of total halo mass at 4 different epochs.  The points are coloured based on the halo spin parameter measured at the same time as the cool disc gas mass.  At each epoch, there is a remarkably tight relationship between cool gas mass and virial mass that does not evolve much between epochs.  No correlation with spin parameter is immediately apparent.

To see how these relationships evolve, the same two lines are drawn in each epoch panel.  At $M_{\rm 200}<2\times10^{11} \msun$, the dotted line represents $M_{gas}\propto M_{\rm 200}^{1.5}$, while at $M_{\rm 200}>2\times10^{11} \msun$, the dashed line shows $M_{\rm gas}=0.04 M_{\rm 200}$.  For $M_{\rm 200}<2\times10^{11} \msun$, the relationship between $M_{\rm 200}$ and $M_{\rm gas}$ evolves with time; it becomes steeper as the galaxies evolve.  At $M_{\rm 200}>2\times10^{11} \msun$, the relationship does not evolve much, but is consistently shallower than $M_{\rm gas}\propto M_{\rm 200}$.

\subsection{The Effect of Hydrodynamics}
\begin{figure*}
\resizebox{18cm}{!}{\includegraphics{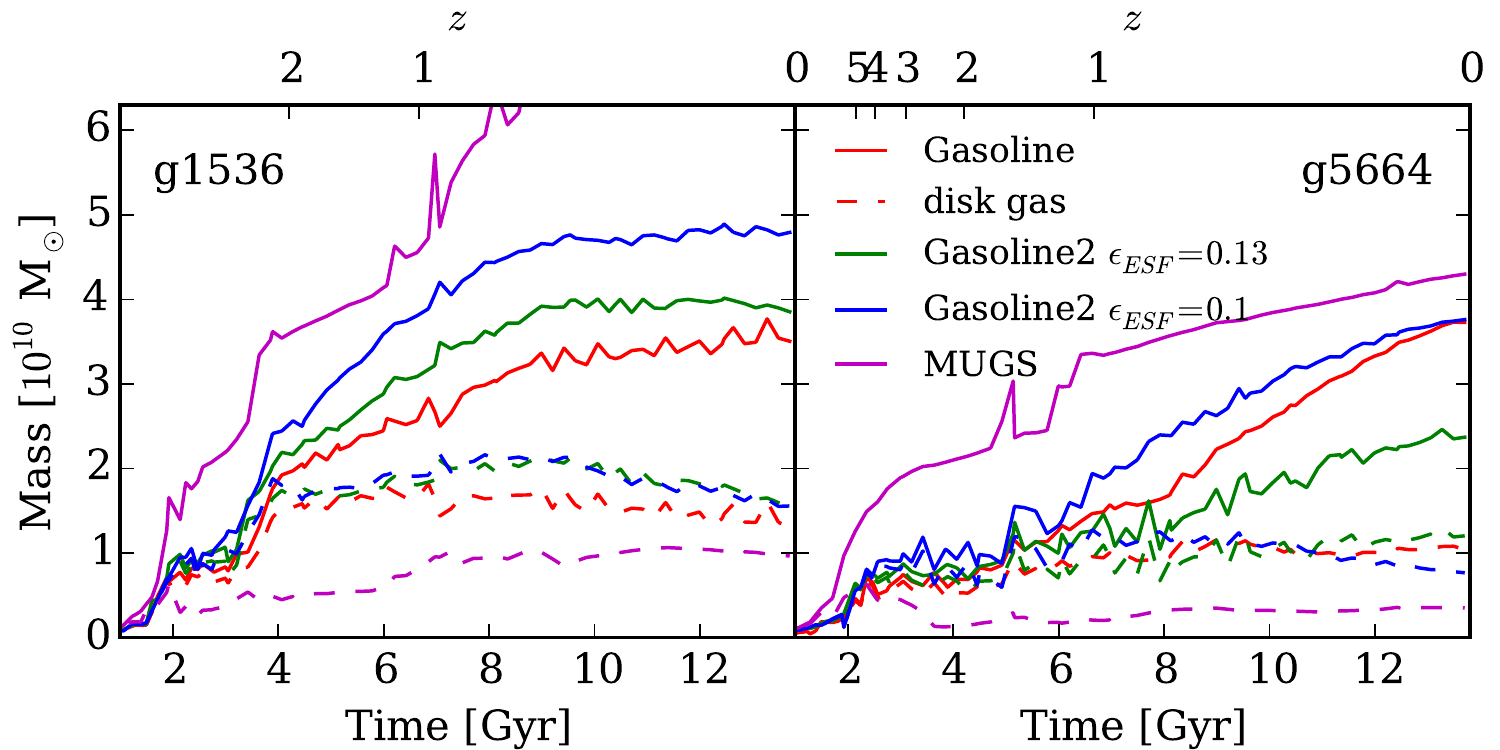}}
 \caption[Disk Mass Evolution]{ The Disc Gas Mass Conspiracy:  A comparison of how the baryonic disc mass evolves using different hydrodynamic models for the galaxies.  The solid line represents all the baroynic mass while the dashed line represents only the gas.  Note that the y-axis is a linear axis in units of $10^{10}$ M$_\odot$.  Cool gas mass is mostly independent of hydrodynamic and feedback treatment.  In all cases, the cool gas mass reaches a nearly constant value.}
\label{fig:discmassev} 
\end{figure*}
While Fig. \ref{fig:lmgas} looks at a sample of galaxies simulated with uniform physics, it is interesting to examine what happens when the physics changes in a limited sample of two simulated galaxies. For this study, we use two cosmological zoom simulations drawn from the McMaster Unbiased Galaxy Simulations \citep[MUGS,][]{Stinson2010}.  The MUGS initial conditions were created using a WMAP3 cosmology \citep[$\Omega_m$=0.24, $\Omega_\Lambda$=0.76, $\Omega_{bary}$=0.044, $H_0=73$ \kms Mpc$^{-1}$, $\sigma_8=0.76$][]{Spergel2007}.  See \citet{Stinson2010} for a complete description of the creation of the initial conditions.  

To study which aspects of the numerical implementation affect the disc gas mass, Fig. \ref{fig:discmassev} shows how the disc gas mass evolves when stellar feedback is varied or when hydrodynamics changes.  The definition for cool gas mass is slightly different in this case.  The gas mass is measured inside a disc region 20 kpc in radius and 6 kpc in height centred on the disc midplane (the disc region includes 3 kpc above and 3 kpc below the midplane) rather than a $\sim40$ kpc sphere.   In Fig. \ref{fig:discmassev}, the solid lines represent the total baryon mass and the dashed lines represent only the gas portion of that mass.

The two galaxies studied here have masses of $\sim7\times10^{11}$ M$_\odot$, but different merger histories.  In both cases, the discs reach a maximum gas mass and remain relatively constant for the rest of the simulation.  In g1536, the change of hydrodynamic scheme from \gas\, to \gastwo\, causes a change in the disc gas mass, but a change in feedback energy in \gastwo\, causes no change in disc gas mass even though the mass of stars nearly doubles in the lower feedback case.  While we show a limited sample here, hydrodynamics likely plays a significant role in determining how much gas mass a disc can support.  Thus, we propose that future studies might use disc gas mass diagnostic for comparing various hydrodynamic schemes.

The largest difference in feedback is shown in the MUGS version of the simulations.  In the MUGS simulations, the feedback was much weaker because MUGS employed a less top-heavy IMF and used only 40\% of the supernova energy for feedback.  Thus, stars output three times less energy per stellar mass in MUGS than the MaGICC simulations.  More than twice as many stars form in MUGS and the MUGS simulations contain less cold gas than any of the others.  In the case of g1536, though MUGS forms so many more stars, the mass of cool gas is only off by $\sim50$\%, and the gas mass stays relatively constant throughout the simulation.  In g5664, the gas mass is more like 30\% the gas mass of the simulations with the other physics, but it again remains largely constant.

\section{Universal Halo Gas Profile}
\label{sec:uni}

Fig. \ref{fig:lmgas} shows that the cool gas mass is well correlated with the halo virial mass.  Fig. \ref{fig:discmassev} shows that a range of feedback treatments on the same initial conditions result in more similar disc gas masses than stellar masses.  With sufficient feedback, the disc gas masses remain constant over many Gyrs.  Combining these facts indicates that the gaseous halos of galaxies might reach some state of hydrostatic equilibrium.  

\subsection{Why is the disc gas mass regulated?}

To understand the steady state gas mass, it may be helpful to think of galaxy formation in the context of the hydrodynamic balance between pressure and gravity \citep[for a more complete description, see \S 4 in][]{White1991}.  The mass of material in a galaxy's halo determines how much gravity is pulling gas into the center of the halo.  The hydrodynamic pressure provides support for the gas.  The balance between these forces configures the gas into something like an isothermal sphere with a density profile of $\rho\propto r^{-2}$.  As the density increases to small radii there is a radius called the cooling radius, $r_c$, inside which the gas cools on a shorter timescale than the dynamical time.  

\citet{Field1969} described the instability of this configuration.  
Gas that cools removes its pressure support, so more gas flows to fill the vacuum, becomes dense and cools.  What can limit this instability is some sort of pressure or energy support that keeps the halo gas at low enough densities so that it does not cool too quickly.  In our simulations, stellar feedback plays this role.  The ultimate fuel source for stellar feedback is the nuclear fusion going on in stars.  However, to create the stellar feedback, stars must be formed, which requires cold gas, which leads to the steady state gas mass.  

\begin{figure}
\resizebox{9cm}{!}{\includegraphics{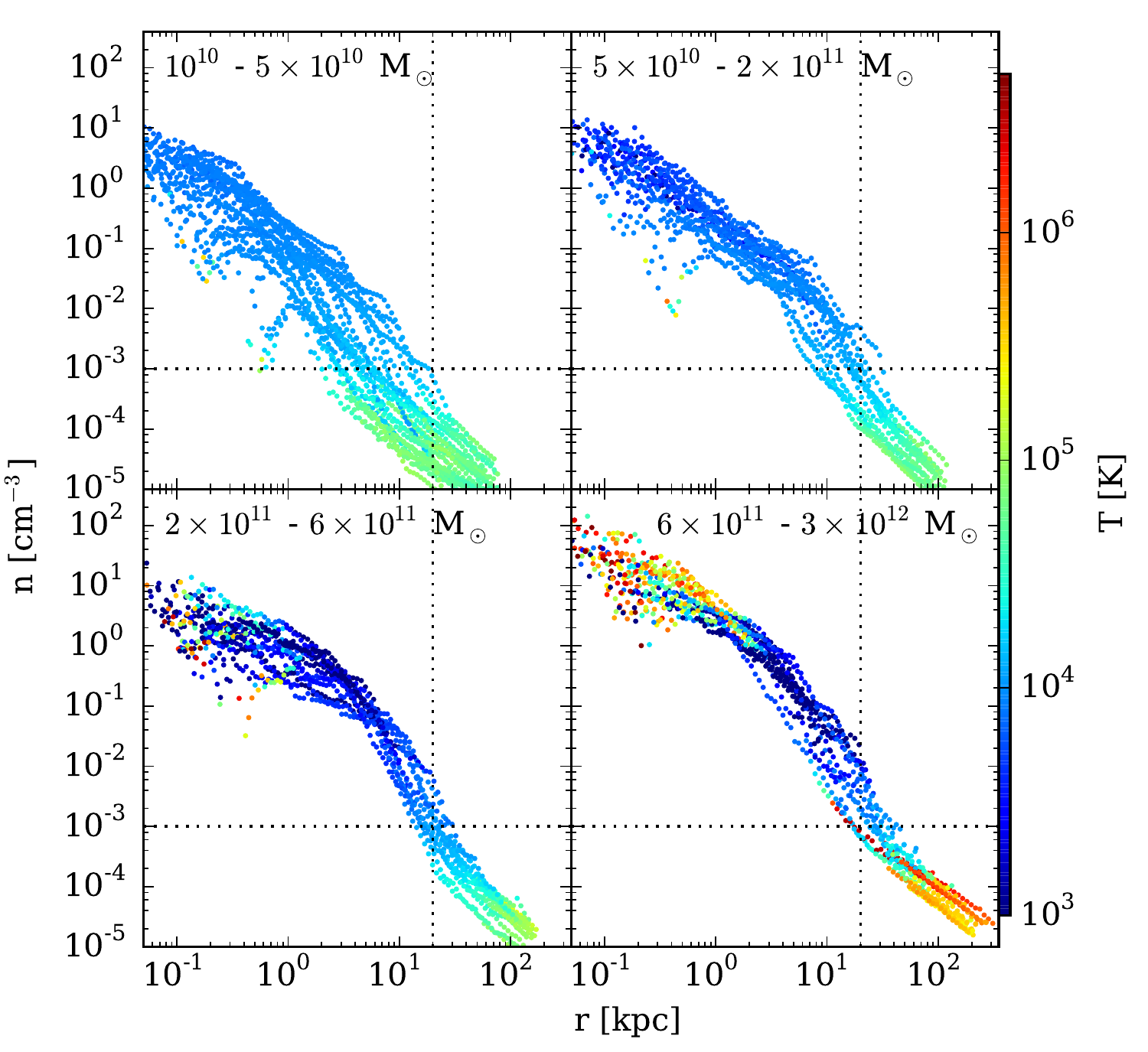}}
 \caption[All Gas Profiles]{ Density profiles for all the gas in the NIHAO sample split into 4 different halo mass ranges.  Each density point is coloured according to the median temperature for the radial bin.  The radii at which the densities are measured are evenly space with the log of the radius.}
\label{fig:allgasprofs} 
\end{figure}

To explore how that equilibrium looks, Fig. \ref{fig:allgasprofs} shows spherically averaged density profiles for all the NIHAO galaxies.  Each point in the profile is coloured according to the median temperature of the gas in that radial bin.  While not all the profiles look exactly the same, they share 3 common sections: the disc, a transition, and an outer halo.  To understand the profile shape, we examine each part sequentially, following the flow of gas from the outside in.  

\begin{figure}
\resizebox{9cm}{!}{\includegraphics{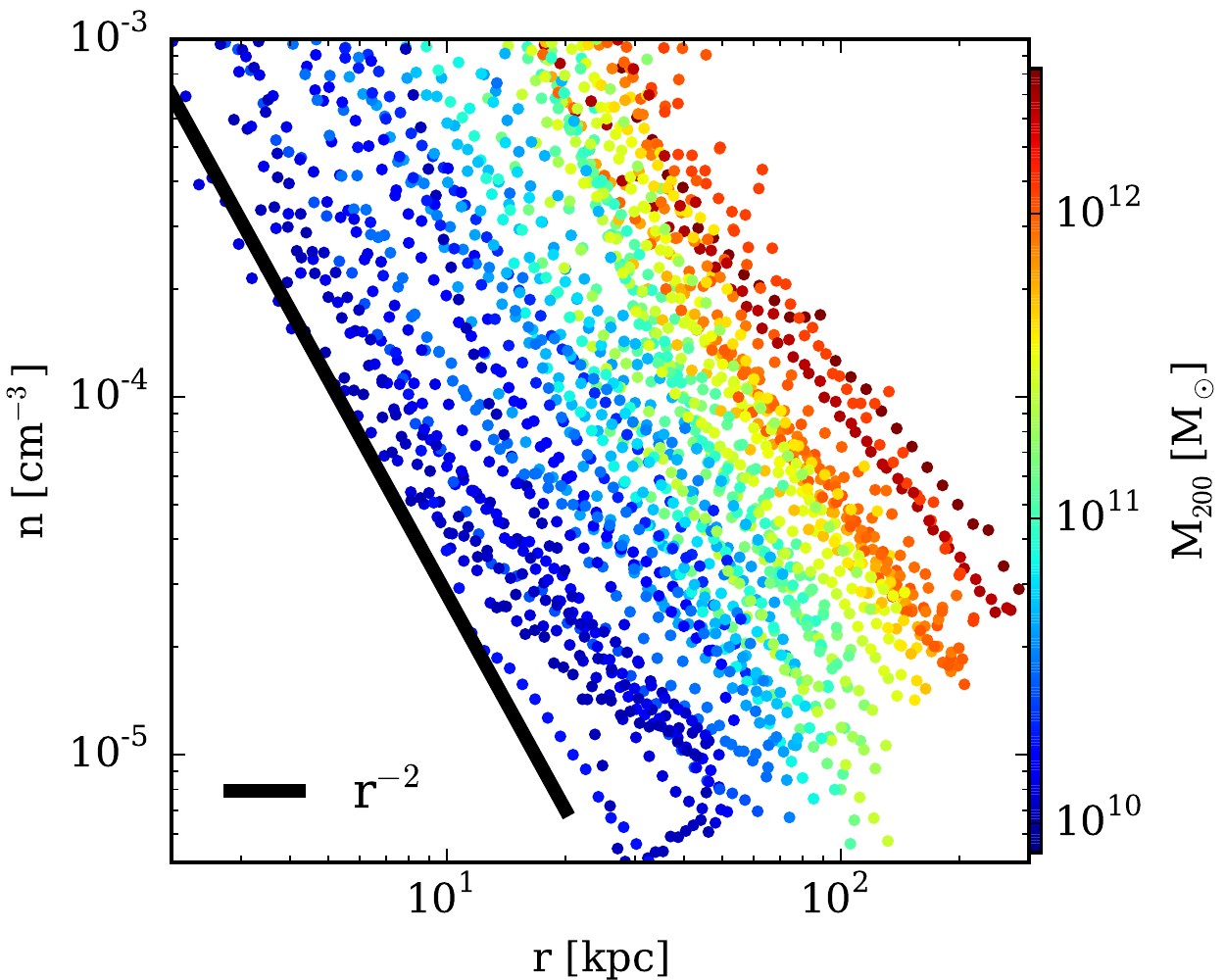}}
 \caption[Outer gas density profile]{ Gas density profile from $r_{200}$ into where the gas density surpasses $10^{-3}$ cm$^{-3}$.  The profiles are coloured according to the the mass (M$_{200}$) of their halo.  There are 100 logarithmically spaced radial bins. }
\label{fig:outer} 
\end{figure}

Fig. \ref{fig:outer} focuses on the outer profile.  The outer halo gas has a nearly isothermal profile with a slope close to -2.  The outer profiles line up neatly according to their halo mass.  Higher mass galaxies have higher virial radii, so the profiles have higher densities at radii more distant from the centre.  All the galaxies in the sample follow a nearly isothermal shape up to a density of $10^{-3}$ cm$^{-3}$.

After the radial profiles reach a density of $10^{-3}$ cm$^{-3}$, the profile transitions to a significantly steeper slope.  Over a short radial range, the density increases dramatically because at this density it is able to cool much more efficiently.  Its cooling time drops below 1 Myr if the temperature is below $10^7$ K.  In the terms described in \citet{White1991}, this is the ``cooling radius.''  

\begin{figure}
\resizebox{9cm}{!}{\includegraphics{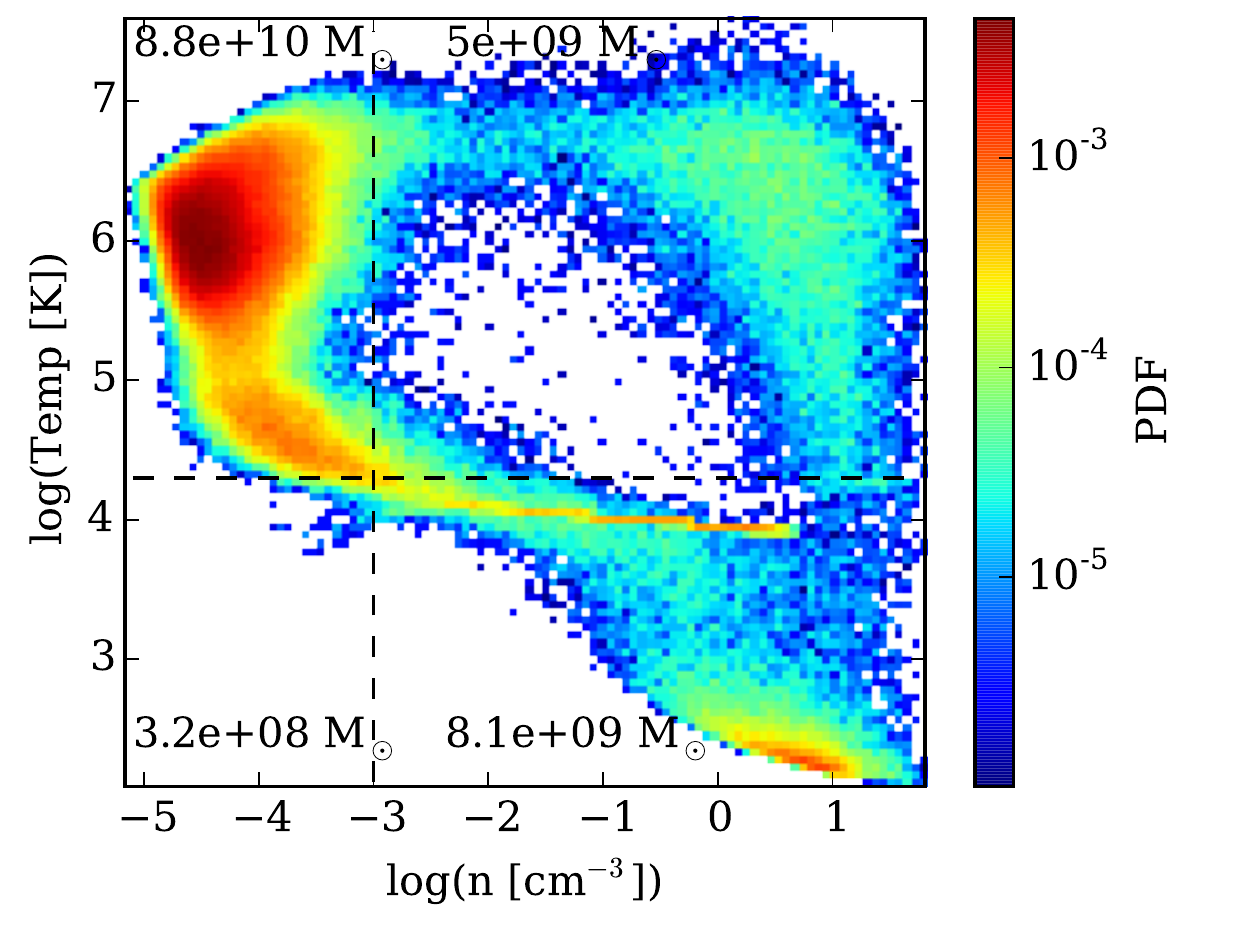}}
 \caption[Gas Phase Diagram]{ Gas temperature-density phase diagram for g1.92e12, a M$_{200}=2.34\times10^{12}$ M$_\odot$ halo. The diagram is split into four parts with the black dashed lines.  The temperature threshold is $2\times10^4$ K and the density is split at $10^{-3}$ cm$^{-3}$.  Each quadrant is labeled with the total gas mass inside it.}
\label{fig:phase} 
\end{figure}
Moving inside the cooling radius, Fig. \ref{fig:phase} illustrates the transition in gas phase that occurs at $n=10^{-3}$ cm$^{-3}$.  At densities below $n<10^{-3}$ cm$^{-3}$, the gas is entirely in a warm ($>20000$ K), diffuse phase.  At higher densities, the gas cools, loses pressure and collapses to reach densities of 0.1-10 cm$^{-3}$.  Images of the cool (T$<$20000 K), dense gas shows that it is in a flattened distribution because it is no longer pressure supported, but instead its centripetal acceleration supports it.  Fig. \ref{fig:allgasprofs} shows that the total gas profile of the disc flattens out to $r^{-1}$ though its surface density profile is exponential.

The inner median temperatures evolve with mass.  Fig. \ref{fig:allgasprofs} shows that lower mass halos have median temperatures of $10^4$ K in their inner regions.  As halo mass increases, the median temperature in the inner region initially drops to $10^3$ K, but in the highest mass halos, the innermost bins reach median temperatures close to $10^6$ K.  

Fig. \ref{fig:phase} is the phase diagram for a high ($3\times10^{12} \msun$) mass halo.  It shows the multiple gas phases present at high densities in the simulations.  Three main phases are obvious.  Below $T<20000$ K, Fig. \ref{fig:phase} shows two phases of gas, one around $10^4$ K and one below $10^3$ K.  The branch below $10^3$ K has higher metallicity.  Higher metallicity gas cools more readily, so as halos grow in mass and form more stars, their gas is enriched with more chemical elements that facilitate cooling to lower temperatures.  At these high masses, the cooling becomes more effective such that star formation increases.  The increase in star formation ejects more energy as feedback that creates more hot, dense gas in these simulations.  

The high mass halos have deeper potential wells, so gas in the centre of the high mass halos exhibit higher median temperatures in Fig. \ref{fig:allgasprofs}.  While gas of $10^6$ K escapes from lower mass halos, $10^{12}$ \msun halos can hold onto this gas.  Fig. \ref{fig:phase} (the phase diagram of a high mass halo) shows that there is almost as much mass in the high temperature, high density phase as cool gas in the disc.  In our simulations, the disabled cooling from stellar feedback artificially maintains gas in this state.  This hot, dense gas is providing the pressure support that prevents a major cooling flow.  At these stellar masses, other observations show that quenching may start to set in.  So, the stellar feedback is less effective than what is happening in the real universe.  Thus, many models employ strong AGN feedback at these mass scales to decrease star formation \citep{Croton2006,Vogelsberger2014}.

One final feature to note in Fig. \ref{fig:allgasprofs} is that lower mass halos exhibit a broad scatter in the density profiles of their inner regions.  The scatter reduces as mass increases.  The scatter comes from stellar feedback that disrupts the gaseous discs.  Lower mass halos have shallower potential wells, so it is easier for the same amount of stellar feedback energy to blow apart the disc.  As the galaxies grow in mass, the stellar feedback has notably less effect.  Fig. \ref{fig:allgasprofs} shows that enough hot gas is created that it becomes the dominant phase in the inner regions of the highest mass galaxies in our sample.

\section{Discussion}
\label{sec:discussion}

\subsection{Theoretical Simplification}
Galaxy formation can seem complex and involves many physical processes.  So, it is comforting when a simple result comes out of a galaxy formation model.  Theorists have noticed that if you assume the mass of cold disc gas remains constant, it is possible to produce simple models of galaxy formation that reproduce many observed properties \citep{Bouche2010,Dave2012,Krumholz2012}.  In these models, sometimes referred to as ``bathtub models'', the amount of gas that is accreted onto galaxies is balanced by sinks of gas, star formation and outflows to leave a constant gas mass.  The reason given for a constant gas mass in these models is that the equations governing the models include the derivative of gas mass with time.  The solution to a simple ordinary differential equation is a decaying exponential, so the time derivative of gas mass should decay to 0 \citep{Dekel2014}.

\subsection{Why Constant Gas Mass?}
\label{sec:discgas}
The inner region of galaxy halos have mean densities corresponding to dynamical times of 100s Myr.  This is less than the age of the Universe, so one would expect that some sort of steady state could be reached.  Specifically, one should compare the dynamical and cooling times with the ``gas depletion timescale.''

The ``gas depletion timescale'' \citep[e.g.][]{Saintonge2011} compares the mass of gas observed in discs with their star formation rate and outflow rate (the gas sinks).  Typical observed timescales are $\sim2$ Gyr, which is an indication of how inefficient star formation is in galaxies: there is a large reservoir of cool gas in galaxy discs ($\sim10^{10}$ M$_\odot$), but only a small fraction of it forms stars ($\sim1$ M$_\odot$ yr$^{-1}$).  

If the cooling and free fall times of halo gas is shorter than the gas depletion timescale, then gas will cool into the disc and be replenished.  Cooling time is a function of temperature and density.  As a galaxy grows, its virial temperature increases.  At some temperature, the gas will be hot enough that its cooling time becomes longer than the gas depletion time.  At this temperature ($\sim10^7$ K), the hot gas will also provide enough pressure to support the hydrostatic equilibrium of the galaxy halo.  This temperature is beyond that which supernova create, but within reach of the temperature gas that outflows from AGN can create.  The gas depletion timescale then sets the time for ``strangulation'' of the galaxy to occur.

\subsection{Comparing M$_{\rm gas}$-M$_{200}$ with M$_\star$-M$_{200}$}
Fig. \ref{fig:lmgas} showed that the relation between M$_{\rm gas}$ and M$_{200}$ has a characteristic shape that is steep, M$_{\rm gas}\sim$M$_{200}^{1.5}$, below M$_{200}=2\times10^{11}\msun$ and then flattens to linear above M$_{200}>2\times10^{11}\msun$.  In comparison, the M$_\star$--M$_{\rm halo}$ relation has a similarly steep slope, M$_\star \propto$ M$_{200}^{1.8}$ for $M_{200}<2\times10^{11} \msun$ halos at $z=0$ \citep{Kravtsov2014}.  

The slopes of M$_{gas}$-M$_{200}$ and M$_\star$-M$_{200}$ also both evolve with redshift.  \citet{Moster2013} find that M$_\star$-M$_{200}$ gets steeper as time evolves, so that it is steepest at z=0.  Fig. \ref{fig:lmgas} shows that the same is true for M$_{gas}$-M$_{200}$.  

Looking above M$_{200}>2\times10^{11}\msun$, there are two cases reported for M$_\star$-M$_{200}$.  \citet{Moster2013} report that M$_\star$-M$_{200}$ becomes shallower, while \citet{Behroozi2013} find M$_\star$--M$_{halo}$ steepens in the range $2\times10^{11} < M_{200} / \msun < 10^{12}$.  The \citet{Moster2013} result indicates that star formation follows the growth of gas mass.  The Behroozi result points to a mass range in which star formation has enhanced efficiency.

\subsection{Observational Verification}
\citet{Prochaska2005} noticed that when you add up all the damped Lyman-$\alpha$ systems seen in quasar absorption lines, it turned out that the \textsc{Hi} density ($\Omega_{HI}$) in the Universe remains constant throughout the history of the Universe.  We do not have a large enough statistical sample to state whether the constant cool disc gas mass corresponds to a constant $\Omega_{HI}$, but it would not be surprising if there is a connection.  Interested readers should see \citet{Lagos2011}, who examines how the cool gas fraction evolves in semi-analytic models.

We have seen in these simulations that the exact treatment of stellar feedback does not greatly affect the steady state disc gas mass.  Stellar feedback determines the efficiency with which accreting gas turns into stars \citep[a la][]{Hopkins2013}.  The disc gas mass is set by the conversion of gas into stars, the Kennicutt-Schmidt relationship \citep{Kennicutt1998,Bigiel2008}.

\citet{White1991} note that the exact disc gas mass is not important for modeling the galaxy luminosity function.  Cold gas is just one phase gas takes as it cools and turns into stars.
However, there are some interesting observational implications.  \citet{Bigiel2012} recently found that isolated galaxies have a universal neutral gas profile that leads them to conclude that neutral gas mass only depends on their optical radius, $r_{25}$.  We are not the only ones seeing a correspondence between neutral gas and simulations, \citet{Wang2014} find that the neutral gas in simulations compares well with observations.

\subsection{Why is $3\times10^{10} \msun$ the maximum gas mass?}
To understand what limits the mass of cool gas in discs, It is instructional to integrate the mass included in an exponential disc with a typical scale radius.  For an exponential profile of the form 
\begin{equation}
\Sigma(R)=\Sigma_0 e^{-\frac{R}{R_d}}
\end{equation}
with a scale length, $R_d$, of  5 kpc, when we integrate out to 5 scale lengths (25 kpc), the mass is $10^{10} \msun$ when $\Sigma_0$, the surface density in the centre, is 100 $\msun$ pc$^{-2}$, nearly the same surface density as typical molecular clouds \citep{Larson1981,McKee2007}.  Molecular clouds are regions where there is sufficient shielding from external UV radiation fields such that molecules will not be dissociated.  Instead, the molecules radiatively cool gas to low enough temperatures to enable collapse to high enough densities for fusion to start and stars to form.  The maximum disc gas mass is the disc mass where star formation becomes efficient.

\subsection{Speculation about quenching star formation}
Since our simulated sample is limited at the high mass end as the simulations overproduce stars in the galactic centres, we are forced to speculate about how star formation is quenched.  Extrapolating our results indicates that increasing virial temperatures possibly combined with other energetic input (AGN?) raise cooling times sufficiently to halt gas accretion onto galaxies and quench star formation.  We see a slight indication of decreasing cool gas mass in our highest mass simulations, but those simulations do not reduce their gas mass quickly enough to correspond to galaxy quenching.

\section{Conclusions}
We study the gas masses in a large set of cosmological hydrodynamic simulations that well reproduce both the halo M$_\star$-M$_{200}$ relation and the M$_\star$-M$_{\rm HI}$ relation in observations.  We find that disc gas masses remain remarkably constant over many Gyrs following their last major merger.  The cool disc masses vary less than 50\% in all galaxies such that most galaxy cool gas masses vary by less than unity over the Hubble time.  Since cool gas masses stay constant for long periods of time, the steady state of galaxies can be modeled using equilibrium or ``bathtub'' models.  

The most important factor in how cool gas mass evolves is how fast the halo spins.  Low spin galaxies decrease their cool disc gas mass while high spin galaxies can add cool gas.  The difference corresponds to the efficiency of star formation.  Low spin galaxies reach higher gas surface densities that lead to more efficient star formation.  Consequently, low spin galaxies have lower disc gas fractions than high spin galaxies.  

Over a wide range of galaxy masses, the constant cool disc gas mass follows a tight correlation with the halo virial mass.  The disc gas mass--halo mass relation has a break at M$_{200}=2\times10^{11} \msun$.  At masses below the break, M$_{\rm gas} \sim$ M$_{200}$; above the break, M$_{\rm gas} \sim$ M$_{200}^{0.5}$.  The low mass relation steepens with time, likely in response to changes in the ionization background.  Observations show a similar break in the M$_{\rm gas}$--M$_\star$ relation, and a comparison shows that the simulations break at a similar stellar mass to the observations.

The tight relation between halo mass and cool gas mass indicates that there is some sort of universal physical mechanism that determines the cool gas mass.  To study the mechanism that sets the cool gas mass, we examine the one-dimensional total gas density profiles of galaxies.  The profiles show three characteristic features across our entire sample.  In the outskirts, the gas halos follow an isothermal profile, $\rho \propto r^{-2}$.  When the density reaches $10^{-3}$ cm$^{-3}$, the density steeply increases to 0.1 cm$^{-3}$ due to the shortened cooling time in dense gas.  Inside that region, the 1D spherical density profile of the disc is a flatter $\rho \propto r^{-1}$ power law, though it shows large scatter in the lowest mass halos.  The scattered density profile results from stellar feedback disrupting the disc in the lowest mass halos.  

The constant gas masses and near universal density profiles lead us to conclude that gas in galaxies reaches a quasi-static equilibrium configuration.  In this configuration, star formation and outflows result from any accretion onto the disc.  Star formation rates and outflows are relatively small due to the large amount of gas mass contained in the diffuse gas halo.  Thus, stellar feedback must be effective at providing pressure support to prevent mass cooling flows from hot halo gas onto the disc for galaxies with lower mass than the Milky Way.  We leave studying how the high entropy (low density, high temperature) gas that stellar feedback produces interacts with the halo for future work.  

In a small experiment that explored how the constant gas mass is affected when various feedbacks are employed, we find that feedback does not have a large effect on the constant gas mass.  While the gas disc masses are relatively constant, the mass of stars in each simulation vary greatly depending on the specifics of the star formation and feedback recipe.  How much cool gas exists in the disc should depend sensitively on the difficult problem of correctly modeling the stark boundary between cold, dense gas in the galaxy disc and the hot, diffuse gas in the halo surrounding galaxies.  Thus, cool gas mass should be a good metric to compare how various hydrodynamic methods simulate galaxies.

\section*{Acknowledgements}
The authors are grateful to Kathryn Kreckel, Eddie Schlafly, Neil Crighton, Hans-Walter Rix, Yuval Birnboim, and Avishai Dekel for useful conversations and comments on the paper.
The analysis was performed using the pynbody package
(\texttt{http://pynbody.github.io}), which was written by Andrew Pontzen and Rok Ro\v{s}kar
in addition to the authors.  
GSS and JH acknowledge funding 
from the European Research Council under the European Union's 
Seventh Framework Programme (FP 7) ERC Grant Agreement n. [321035].
The simulations were performed on the \textsc{theo} and \textsc{hydra} clusters of  the
Max-Planck-Institut f\"ur Astronomie at the Rechenzentrum in Garching;
the clusters hosted on \textsc{sharcnet}, part of ComputeCanada.  We greatly appreciate
the contributions of these computing allocations.
CBB acknowledges Max- Planck-Institut f\"ur Astronomie for its hospitality and financial support through the  Sonderforschungsbereich SFB 881 ``The Milky Way System''
(subproject A1) of the German Research Foundation (DFG).  AVM, AAD and GS also acknowledge support from SFB 881 (subproject A1) of the DFG.  JW thanks NSERC for support.  TQ acknowledges support from grants AST-0908499 and AST-1311956 from the NSF. 
\bibliographystyle{mn2e}
\bibliography{references}

\begin{thebibliography}{}

\bibitem[\protect\citeauthoryear{{Aumer}, {White}, {Naab} \&
  {Scannapieco}}{{Aumer} et~al.}{2013}]{Aumer2013}
{Aumer} M.,  {White} S.~D.~M.,  {Naab} T.,    {Scannapieco} C.,  2013, \mnras,
  434, 3142

\bibitem[\protect\citeauthoryear{{Bauermeister}, {Blitz} \&
  {Ma}}{{Bauermeister} et~al.}{2010}]{Bauermeister2010}
{Bauermeister} A.,  {Blitz} L.,    {Ma} C.-P.,  2010, \apj, 717, 323

\bibitem[\protect\citeauthoryear{{Behroozi}, {Wechsler} \& {Conroy}}{{Behroozi}
  et~al.}{2013}]{Behroozi2013}
{Behroozi} P.~S.,  {Wechsler} R.~H.,    {Conroy} C.,  2013, \apj, 770, 57

\bibitem[\protect\citeauthoryear{{Bigiel} \& {Blitz}}{{Bigiel} \&
  {Blitz}}{2012}]{Bigiel2012}
{Bigiel} F.,  {Blitz} L.,  2012, \apj, 756, 183

\bibitem[\protect\citeauthoryear{{Bigiel}, {Leroy}, {Walter}, {Brinks}, {de
  Blok}, {Madore} \& {Thornley}}{{Bigiel} et~al.}{2008}]{Bigiel2008}
{Bigiel} F.,  {Leroy} A.,  {Walter} F.,  {Brinks} E.,  {de Blok} W.~J.~G.,
  {Madore} B.,    {Thornley} M.~D.,  2008, \aj, 136, 2846

\bibitem[\protect\citeauthoryear{{Blitz} \& {Rosolowsky}}{{Blitz} \&
  {Rosolowsky}}{2006}]{Blitz2006}
{Blitz} L.,  {Rosolowsky} E.,  2006, \apj, 650, 933

\bibitem[\protect\citeauthoryear{{Blyth}, {van der Hulst}, {Verheijen}, {SWG
  Members}, {Catinella}, {Fraternali}, {Haynes}, {Hess}, {Koribalski}, {Lagos},
  {Meyer}, {Obreschkow}, {Popping}, {Power}, {Verdes-Montenegro} \&
  {Zwaan}}{{Blyth} et~al.}{2015}]{SKA}
{Blyth} S.-L.,  {van der Hulst} J.~M.,  {Verheijen} M.~A.~W.,  {SWG Members}
  H.,  {Catinella} B.,  {Fraternali} F.,  {Haynes} M.~P.,  {Hess} K.~M.,
  {Koribalski} B.~S.,  {Lagos} C.,  {Meyer} M.,  {Obreschkow} D.,  {Popping}
  A.,  {Power} C.,  {Verdes-Montenegro} L.,    {Zwaan} M.,  2015, ArXiv
  e-prints

\bibitem[\protect\citeauthoryear{{Bouch{\'e}}, {Dekel}, {Genzel}, {Genel},
  {Cresci}, {F{\"o}rster Schreiber}, {Shapiro}, {Davies} \&
  {Tacconi}}{{Bouch{\'e}} et~al.}{2010}]{Bouche2010}
{Bouch{\'e}} N.,  {Dekel} A.,  {Genzel} R.,  {Genel} S.,  {Cresci} G.,
  {F{\"o}rster Schreiber} N.~M.,  {Shapiro} K.~L.,  {Davies} R.~I.,
  {Tacconi} L.,  2010, \apj, 718, 1001

\bibitem[\protect\citeauthoryear{{Bradford}, {Geha} \& {Blanton}}{{Bradford}
  et~al.}{2015}]{Bradford2015}
{Bradford} J.~D.,  {Geha} M.~C.,    {Blanton} M.~R.,  2015, ArXiv e-prints

\bibitem[\protect\citeauthoryear{{Brook}, {Stinson}, {Gibson}, {Ro{\v s}kar},
  {Wadsley} \& {Quinn}}{{Brook} et~al.}{012a}]{Brook2012}
{Brook} C.~B.,  {Stinson} G.,  {Gibson} B.~K.,  {Ro{\v s}kar} R.,  {Wadsley}
  J.,    {Quinn} T.,  2012a, \mnras, 419, 771

\bibitem[\protect\citeauthoryear{{Brook}, {Stinson}, {Gibson}, {Shen},
  {Macci{\`o}}, {Obreja}, {Wadsley} \& {Quinn}}{{Brook}
  et~al.}{2014}]{Brook2014}
{Brook} C.~B.,  {Stinson} G.,  {Gibson} B.~K.,  {Shen} S.,  {Macci{\`o}} A.~V.,
   {Obreja} A.,  {Wadsley} J.,    {Quinn} T.,  2014, \mnras, 443, 3809

\bibitem[\protect\citeauthoryear{{Bullock}, {Dekel}, {Kolatt}, {Kravtsov},
  {Klypin}, {Porciani} \& {Primack}}{{Bullock} et~al.}{2001}]{Bullock2001}
{Bullock} J.~S.,  {Dekel} A.,  {Kolatt} T.~S.,  {Kravtsov} A.~V.,  {Klypin}
  A.~A.,  {Porciani} C.,    {Primack} J.~R.,  2001, \apj, 555, 240

\bibitem[\protect\citeauthoryear{{Catinella}, {Schiminovich}, {Kauffmann},
  {Fabello}, {Hummels}, {Lemonias}, {Moran}, {Wu}, {Cooper} \&
  {Wang}}{{Catinella} et~al.}{2012}]{Catinella2012}
{Catinella} B.,  {Schiminovich} D.,  {Kauffmann} G.,  {Fabello} S.,  {Hummels}
  C.,  {Lemonias} J.,  {Moran} S.~M.,  {Wu} R.,  {Cooper} A.,    {Wang} J.,
  2012, \aap, 544, A65

\bibitem[\protect\citeauthoryear{{Chabrier}}{{Chabrier}}{2003}]{Chabrier2003}
{Chabrier} G.,  2003, \pasp, 115, 763

\bibitem[\protect\citeauthoryear{{Crighton}, {Murphy}, {Prochaska}, {Worseck},
  {Rafelski}, {Becker}, {Ellison}, {Fumagalli}, {Lopez}, {Meiksin} \&
  {O'Meara}}{{Crighton} et~al.}{2015}]{Crighton2015}
{Crighton} N.~H.~M.,  {Murphy} M.~T.,  {Prochaska} J.~X.,  {Worseck} G.,
  {Rafelski} M.,  {Becker} G.~D.,  {Ellison} S.~L.,  {Fumagalli} M.,  {Lopez}
  S.,  {Meiksin} A.,    {O'Meara} J.~M.,  2015, ArXiv e-prints

\bibitem[\protect\citeauthoryear{{Croton}, {Springel}, {White}, {De Lucia},
  {Frenk}, {Gao}, {Jenkins}, {Kauffmann}, {Navarro} \& {Yoshida}}{{Croton}
  et~al.}{2006}]{Croton2006}
{Croton} D.~J.,  {Springel} V.,  {White} S.~D.~M.,  {De Lucia} G.,  {Frenk}
  C.~S.,  {Gao} L.,  {Jenkins} A.,  {Kauffmann} G.,  {Navarro} J.~F.,
  {Yoshida} N.,  2006, \mnras, 365, 11

\bibitem[\protect\citeauthoryear{{Daddi}, {Bournaud}, {Walter}, {Dannerbauer},
  {Carilli}, {Dickinson}, {Elbaz}, {Morrison}, {Riechers}, {Onodera}, {Salmi},
  {Krips} \& {Stern}}{{Daddi} et~al.}{2010}]{Daddi2010}
{Daddi} E.,  {Bournaud} F.,  {Walter} F.,  {Dannerbauer} H.,  {Carilli} C.~L.,
  {Dickinson} M.,  {Elbaz} D.,  {Morrison} G.~E.,  {Riechers} D.,  {Onodera}
  M.,  {Salmi} F.,  {Krips} M.,    {Stern} D.,  2010, \apj, 713, 686

\bibitem[\protect\citeauthoryear{{Dav{\'e}}, {Finlator} \&
  {Oppenheimer}}{{Dav{\'e}} et~al.}{2012}]{Dave2012}
{Dav{\'e}} R.,  {Finlator} K.,    {Oppenheimer} B.~D.,  2012, \mnras, 421, 98

\bibitem[\protect\citeauthoryear{{Dehnen} \& {Aly}}{{Dehnen} \&
  {Aly}}{2012}]{Dehnen2012}
{Dehnen} W.,  {Aly} H.,  2012, \mnras, 425, 1068

\bibitem[\protect\citeauthoryear{{Dekel} \& {Mandelker}}{{Dekel} \&
  {Mandelker}}{2014}]{Dekel2014}
{Dekel} A.,  {Mandelker} N.,  2014, \mnras, 444, 2071

\bibitem[\protect\citeauthoryear{{Dekel}, {Zolotov}, {Tweed}, {Cacciato},
  {Ceverino} \& {Primack}}{{Dekel} et~al.}{2013}]{Dekel2013}
{Dekel} A.,  {Zolotov} A.,  {Tweed} D.,  {Cacciato} M.,  {Ceverino} D.,
  {Primack} J.~R.,  2013, \mnras, 435, 999

\bibitem[\protect\citeauthoryear{{Dutton} \& {Macci{\`o}}}{{Dutton} \&
  {Macci{\`o}}}{2014}]{Dutton2014}
{Dutton} A.~A.,  {Macci{\`o}} A.~V.,  2014, \mnras, 441, 3359

\bibitem[\protect\citeauthoryear{{Dutton}, {van den Bosch} \& {Dekel}}{{Dutton}
  et~al.}{2010}]{Dutton2010}
{Dutton} A.~A.,  {van den Bosch} F.~C.,    {Dekel} A.,  2010, \mnras, 405, 1690

\bibitem[\protect\citeauthoryear{{Ferland}, {Korista}, {Verner}, {Ferguson},
  {Kingdon} \& {Verner}}{{Ferland} et~al.}{1998}]{Ferland1998}
{Ferland} G.~J.,  {Korista} K.~T.,  {Verner} D.~A.,  {Ferguson} J.~W.,
  {Kingdon} J.~B.,    {Verner} E.~M.,  1998, \pasp, 110, 761

\bibitem[\protect\citeauthoryear{{Field}, {Goldsmith} \& {Habing}}{{Field}
  et~al.}{1969}]{Field1969}
{Field} G.~B.,  {Goldsmith} D.~W.,    {Habing} H.~J.,  1969, \apjl, 155, L149

\bibitem[\protect\citeauthoryear{{Finlator} \& {Dav{\'e}}}{{Finlator} \&
  {Dav{\'e}}}{2008}]{Finlator2008}
{Finlator} K.,  {Dav{\'e}} R.,  2008, \mnras, 385, 2181

\bibitem[\protect\citeauthoryear{{Fraternali} \& {Binney}}{{Fraternali} \&
  {Binney}}{2008}]{Fraternali2008}
{Fraternali} F.,  {Binney} J.~J.,  2008, \mnras, 386, 935

\bibitem[\protect\citeauthoryear{{Genel}, {Vogelsberger}, {Springel},
  {Sijacki}, {Nelson}, {Snyder}, {Rodriguez-Gomez}, {Torrey} \&
  {Hernquist}}{{Genel} et~al.}{2014}]{Genel2014}
{Genel} S.,  {Vogelsberger} M.,  {Springel} V.,  {Sijacki} D.,  {Nelson} D.,
  {Snyder} G.,  {Rodriguez-Gomez} V.,  {Torrey} P.,    {Hernquist} L.,  2014,
  \mnras, 445, 175

\bibitem[\protect\citeauthoryear{{Genzel}, {Tacconi}, {Gracia-Carpio},
  {Sternberg}, {Cooper}, {Shapiro}, {Bolatto}, {Bouch{\'e}}, {Bournaud},
  {Burkert}, {Combes}, {Comerford}, {Cox} \& {Davis}}{{Genzel}
  et~al.}{2010}]{Genzel2010}
{Genzel} R.,  {Tacconi} L.~J.,  {Gracia-Carpio} J.,  {Sternberg} A.,  {Cooper}
  M.~C.,  {Shapiro} K.,  {Bolatto} A.,  {Bouch{\'e}} N.,  {Bournaud} F.,
  {Burkert} A.,  {Combes} F.,  {Comerford} J.,  {Cox} P.,    {Davis} M.,  2010,
  \mnras, 407, 2091

\bibitem[\protect\citeauthoryear{{Giovanelli}, {Haynes}, {Kent}, {Perillat},
  {Saintonge}, {Brosch}, {Catinella}, {Hoffman}, {Stierwalt}, {Spekkens},
  {Lerner}, {Masters} \& {Momjian}}{{Giovanelli} et~al.}{2005}]{Giovanelli2005}
{Giovanelli} R.,  {Haynes} M.~P.,  {Kent} B.~R.,  {Perillat} P.,  {Saintonge}
  A.,  {Brosch} N.,  {Catinella} B.,  {Hoffman} G.~L.,  {Stierwalt} S.,
  {Spekkens} K.,  {Lerner} M.~S.,  {Masters} K.~L.,    {Momjian} E.,  2005,
  \aj, 130, 2598

\bibitem[\protect\citeauthoryear{{Haardt} \& {Madau}}{{Haardt} \&
  {Madau}}{2005}]{Haardt2005}
{Haardt} F.,  {Madau} P.,  2005, unpublished

\bibitem[\protect\citeauthoryear{{Haynes}, {Giovanelli}, {Martin}, {Hess},
  {Saintonge}, {Adams}, {Hallenbeck}, {Hoffman}, {Huang}, {Kent}, {Koopmann},
  {Papastergis}, {Stierwalt}, {Balonek}, {Craig}, {Higdon} \&
  {Kornreich}}{{Haynes} et~al.}{2011}]{Haynes2011}
{Haynes} M.~P.,  {Giovanelli} R.,  {Martin} A.~M.,  {Hess} K.~M.,  {Saintonge}
  A.,  {Adams} E.~A.~K.,  {Hallenbeck} G.,  {Hoffman} G.~L.,  {Huang} S.,
  {Kent} B.~R.,  {Koopmann} R.~A.,  {Papastergis} E.,  {Stierwalt} S.,
  {Balonek} T.~J.,  {Craig} D.~W.,  {Higdon} S.~J.~U.,    {Kornreich} D.~A.,
  2011, \aj, 142, 170

\bibitem[\protect\citeauthoryear{{Herpich}, {Stinson}, {Dutton}, {Rix},
  {Martig}, {Ro{\v s}kar}, {Macci{\`o}}, {Quinn} \& {Wadsley}}{{Herpich}
  et~al.}{2015}]{Herpich2015}
{Herpich} J.,  {Stinson} G.~S.,  {Dutton} A.~A.,  {Rix} H.-W.,  {Martig} M.,
  {Ro{\v s}kar} R.,  {Macci{\`o}} A.~V.,  {Quinn} T.~R.,    {Wadsley} J.,
  2015, \mnras, 448, L99

\bibitem[\protect\citeauthoryear{{Hopkins}}{{Hopkins}}{2013}]{Hopkins2013}
{Hopkins} P.~F.,  2013, \mnras, 428, 2840

\bibitem[\protect\citeauthoryear{{Hopkins}, {Kere{\v s}}, {O{\~n}orbe},
  {Faucher-Gigu{\`e}re}, {Quataert}, {Murray} \& {Bullock}}{{Hopkins}
  et~al.}{2014}]{Hopkins2014}
{Hopkins} P.~F.,  {Kere{\v s}} D.,  {O{\~n}orbe} J.,  {Faucher-Gigu{\`e}re}
  C.-A.,  {Quataert} E.,  {Murray} N.,    {Bullock} J.~S.,  2014, \mnras, 445,
  581

\bibitem[\protect\citeauthoryear{{Huang}, {Haynes}, {Giovanelli} \&
  {Brinchmann}}{{Huang} et~al.}{2012}]{Huang2012}
{Huang} S.,  {Haynes} M.~P.,  {Giovanelli} R.,    {Brinchmann} J.,  2012, \apj,
  756, 113

\bibitem[\protect\citeauthoryear{{Kalirai}, {Hansen}, {Kelson}, {Reitzel},
  {Rich} \& {Richer}}{{Kalirai} et~al.}{2008}]{Kalirai2008}
{Kalirai} J.~S.,  {Hansen} B.~M.~S.,  {Kelson} D.~D.,  {Reitzel} D.~B.,  {Rich}
  R.~M.,    {Richer} H.~B.,  2008, \apj, 676, 594

\bibitem[\protect\citeauthoryear{{Kannan}, {Stinson}, {Macci{\`o}}, {Hennawi},
  {Woods}, {Wadsley}, {Shen}, {Robitaille}, {Cantalupo}, {Quinn} \&
  {Christensen}}{{Kannan} et~al.}{2014}]{Kannan2014}
{Kannan} R.,  {Stinson} G.~S.,  {Macci{\`o}} A.~V.,  {Hennawi} J.~F.,  {Woods}
  R.,  {Wadsley} J.,  {Shen} S.,  {Robitaille} T.,  {Cantalupo} S.,  {Quinn}
  T.~R.,    {Christensen} C.,  2014, \mnras, 437, 2882

\bibitem[\protect\citeauthoryear{{Keller}, {Wadsley}, {Benincasa} \&
  {Couchman}}{{Keller} et~al.}{2014}]{Keller2014}
{Keller} B.~W.,  {Wadsley} J.,  {Benincasa} S.~M.,    {Couchman} H.~M.~P.,
  2014, \mnras, 442, 3013

\bibitem[\protect\citeauthoryear{{Kennicutt}}{{Kennicutt}}{1998}]{Kennicutt1998}
{Kennicutt} R.~C.,  1998, \apj, 498, 541+

\bibitem[\protect\citeauthoryear{{Kere{\v s}}, {Katz}, {Weinberg} \&
  {Dav{\'e}}}{{Kere{\v s}} et~al.}{2005}]{Keres2005}
{Kere{\v s}} D.,  {Katz} N.,  {Weinberg} D.~H.,    {Dav{\'e}} R.,  2005,
  \mnras, 363, 2

\bibitem[\protect\citeauthoryear{{Kravtsov}, {Vikhlinin} \&
  {Meshscheryakov}}{{Kravtsov} et~al.}{2014}]{Kravtsov2014}
{Kravtsov} A.,  {Vikhlinin} A.,    {Meshscheryakov} A.,  2014, ArXiv e-prints
  1401.7329

\bibitem[\protect\citeauthoryear{{Krumholz} \& {Dekel}}{{Krumholz} \&
  {Dekel}}{2012}]{Krumholz2012}
{Krumholz} M.~R.,  {Dekel} A.,  2012, \apj, 753, 16

\bibitem[\protect\citeauthoryear{{Lagos}, {Baugh}, {Lacey}, {Benson}, {Kim} \&
  {Power}}{{Lagos} et~al.}{2011}]{Lagos2011}
{Lagos} C.~D.~P.,  {Baugh} C.~M.,  {Lacey} C.~G.,  {Benson} A.~J.,  {Kim}
  H.-S.,    {Power} C.,  2011, \mnras, 418, 1649

\bibitem[\protect\citeauthoryear{{Larson}}{{Larson}}{1981}]{Larson1981}
{Larson} R.~B.,  1981, \mnras, 194, 809

\bibitem[\protect\citeauthoryear{{Leitner} \& {Kravtsov}}{{Leitner} \&
  {Kravtsov}}{2011}]{Leitner2011}
{Leitner} S.~N.,  {Kravtsov} A.~V.,  2011, \apj, 734, 48

\bibitem[\protect\citeauthoryear{{Lilly}, {Carollo}, {Pipino}, {Renzini} \&
  {Peng}}{{Lilly} et~al.}{2013}]{Lilly2013}
{Lilly} S.~J.,  {Carollo} C.~M.,  {Pipino} A.,  {Renzini} A.,    {Peng} Y.,
  2013, \apj, 772, 119

\bibitem[\protect\citeauthoryear{{Lu}, {Mo} \& {Lu}}{{Lu}
  et~al.}{2015}]{Lu2014}
{Lu} Z.,  {Mo} H.~J.,    {Lu} Y.,  2015, \mnras, 450, 606

\bibitem[\protect\citeauthoryear{{Macci{\`o}}, {Dutton} \& {van den
  Bosch}}{{Macci{\`o}} et~al.}{2008}]{Maccio2008}
{Macci{\`o}} A.~V.,  {Dutton} A.~A.,    {van den Bosch} F.~C.,  2008, \mnras,
  391, 1940

\bibitem[\protect\citeauthoryear{{Marinacci}, {Binney}, {Fraternali}, {Nipoti},
  {Ciotti} \& {Londrillo}}{{Marinacci} et~al.}{2010}]{Marinacci2010}
{Marinacci} F.,  {Binney} J.,  {Fraternali} F.,  {Nipoti} C.,  {Ciotti} L.,
  {Londrillo} P.,  2010, \mnras, 404, 1464

\bibitem[\protect\citeauthoryear{{McKee} \& {Ostriker}}{{McKee} \&
  {Ostriker}}{2007}]{McKee2007}
{McKee} C.~F.,  {Ostriker} E.~C.,  2007, \araa, 45, 565

\bibitem[\protect\citeauthoryear{{Meyer}, {Zwaan}, {Webster}, {Staveley-Smith},
  {Ryan-Weber}, {Drinkwater}, {Barnes}, {Howlett}, {Kilborn}, {Stevens},
  {Waugh}, {Pierce}, {Bhathal}, {de Blok}, {Disney} \& {Ekers}}{{Meyer}
  et~al.}{2004}]{Meyer2004}
{Meyer} M.~J.,  {Zwaan} M.~A.,  {Webster} R.~L.,  {Staveley-Smith} L.,
  {Ryan-Weber} E.,  {Drinkwater} M.~J.,  {Barnes} D.~G.,  {Howlett} M.,
  {Kilborn} V.~A.,  {Stevens} J.,  {Waugh} M.,  {Pierce} M.~J.,  {Bhathal} R.,
  {de Blok} W.~J.~G.,  {Disney} M.~J.,    {Ekers} R.~D. e.~a.,  2004, \mnras,
  350, 1195

\bibitem[\protect\citeauthoryear{{Mo}, {Mao} \& {White}}{{Mo}
  et~al.}{1998}]{Mo1998}
{Mo} H.~J.,  {Mao} S.,    {White} S.~D.~M.,  1998, \mnras, 295, 319

\bibitem[\protect\citeauthoryear{{Moster}, {Naab} \& {White}}{{Moster}
  et~al.}{2013}]{Moster2013}
{Moster} B.~P.,  {Naab} T.,    {White} S.~D.~M.,  2013, \mnras, 428, 3121

\bibitem[\protect\citeauthoryear{{Nelson}, {Genel}, {Pillepich},
  {Vogelsberger}, {Springel} \& {Hernquist}}{{Nelson}
  et~al.}{2015}]{Nelson2015}
{Nelson} D.,  {Genel} S.,  {Pillepich} A.,  {Vogelsberger} M.,  {Springel} V.,
    {Hernquist} L.,  2015, ArXiv e-prints

\bibitem[\protect\citeauthoryear{{Oppenheimer}, {Dav{\'e}}, {Kere{\v s}},
  {Fardal}, {Katz}, {Kollmeier} \& {Weinberg}}{{Oppenheimer}
  et~al.}{2010}]{Oppenheimer2010}
{Oppenheimer} B.~D.,  {Dav{\'e}} R.,  {Kere{\v s}} D.,  {Fardal} M.,  {Katz}
  N.,  {Kollmeier} J.~A.,    {Weinberg} D.~H.,  2010, \mnras, 406, 2325

\bibitem[\protect\citeauthoryear{{Planck Collaboration}, {Ade}, {Aghanim},
  {Armitage-Caplan}, {Arnaud}, {Ashdown}, {Atrio-Barandela}, {Aumont},
  {Baccigalupi}, {Banday} \& et al.}{{Planck Collaboration}
  et~al.}{2014}]{Planck2014}
{Planck Collaboration} {Ade} P.~A.~R.,  {Aghanim} N.,  {Armitage-Caplan} C.,
  {Arnaud} M.,  {Ashdown} M.,  {Atrio-Barandela} F.,  {Aumont} J.,
  {Baccigalupi} C.,  {Banday} A.~J.,    et al. 2014, \aap, 571, A16

\bibitem[\protect\citeauthoryear{{Popping}, {Behroozi} \& {Peeples}}{{Popping}
  et~al.}{2015}]{Popping2015}
{Popping} G.,  {Behroozi} P.~S.,    {Peeples} M.~S.,  2015, \mnras, 449, 477

\bibitem[\protect\citeauthoryear{{Popping}, {Somerville} \& {Trager}}{{Popping}
  et~al.}{2014}]{Popping2014}
{Popping} G.,  {Somerville} R.~S.,    {Trager} S.~C.,  2014, \mnras, 442, 2398

\bibitem[\protect\citeauthoryear{{Price}}{{Price}}{2008}]{Price2008}
{Price} D.~J.,  2008, Journal of Computational Physics, 227, 10040

\bibitem[\protect\citeauthoryear{{Prochaska}, {Herbert-Fort} \&
  {Wolfe}}{{Prochaska} et~al.}{2005}]{Prochaska2005}
{Prochaska} J.~X.,  {Herbert-Fort} S.,    {Wolfe} A.~M.,  2005, \apj, 635, 123

\bibitem[\protect\citeauthoryear{{Putman}, {Peek} \& {Joung}}{{Putman}
  et~al.}{2012}]{Putman2012}
{Putman} M.~E.,  {Peek} J.~E.~G.,    {Joung} M.~R.,  2012, \araa, 50, 491

\bibitem[\protect\citeauthoryear{{Rees} \& {Ostriker}}{{Rees} \&
  {Ostriker}}{1977}]{Rees1977}
{Rees} M.~J.,  {Ostriker} J.~P.,  1977, \mnras, 179, 541

\bibitem[\protect\citeauthoryear{{Ritchie} \& {Thomas}}{{Ritchie} \&
  {Thomas}}{2001}]{Ritchie2001}
{Ritchie} B.~W.,  {Thomas} P.~A.,  2001, \mnras, 323, 743

\bibitem[\protect\citeauthoryear{{Saintonge}, {Kauffmann}, {Wang}, {Kramer},
  {Tacconi}, {Buchbender}, {Catinella}, {Graci{\'a}-Carpio}, {Cortese},
  {Fabello}, {Fu}, {Genzel}, {Giovanelli}, {Guo} \& {Haynes}}{{Saintonge}
  et~al.}{2011}]{Saintonge2011}
{Saintonge} A.,  {Kauffmann} G.,  {Wang} J.,  {Kramer} C.,  {Tacconi} L.~J.,
  {Buchbender} C.,  {Catinella} B.,  {Graci{\'a}-Carpio} J.,  {Cortese} L.,
  {Fabello} S.,  {Fu} J.,  {Genzel} R.,  {Giovanelli} R.,  {Guo} Q.,
  {Haynes} M.~P.,  2011, \mnras, 415, 61

\bibitem[\protect\citeauthoryear{{Saitoh} \& {Makino}}{{Saitoh} \&
  {Makino}}{2009}]{Saitoh2009}
{Saitoh} T.~R.,  {Makino} J.,  2009, \apjl, 697, L99

\bibitem[\protect\citeauthoryear{{Schaye}, {Crain}, {Bower}, {Furlong},
  {Schaller}, {Theuns}, {Dalla Vecchia}, {Frenk}, {McCarthy}, {Helly},
  {Jenkins}, {Rosas-Guevara}, {White}, {Baes}, {Booth} \& {Camps}}{{Schaye}
  et~al.}{2015}]{Schaye2015}
{Schaye} J.,  {Crain} R.~A.,  {Bower} R.~G.,  {Furlong} M.,  {Schaller} M.,
  {Theuns} T.,  {Dalla Vecchia} C.,  {Frenk} C.~S.,  {McCarthy} I.~G.,  {Helly}
  J.~C.,  {Jenkins} A.,  {Rosas-Guevara} Y.~M.,  {White} S.~D.~M.,  {Baes} M.,
  {Booth} C.~M.,    {Camps} P.,  2015, \mnras, 446, 521

\bibitem[\protect\citeauthoryear{{Schmidt}}{{Schmidt}}{1959}]{Schmidt1959}
{Schmidt} M.,  1959, \apj, 129, 243

\bibitem[\protect\citeauthoryear{{Shen}, {Wadsley} \& {Stinson}}{{Shen}
  et~al.}{2010}]{Shen2010}
{Shen} S.,  {Wadsley} J.,    {Stinson} G.,  2010, \mnras, 407, 1581

\bibitem[\protect\citeauthoryear{{Somerville} \& {Primack}}{{Somerville} \&
  {Primack}}{1999}]{Somerville1999}
{Somerville} R.~S.,  {Primack} J.~R.,  1999, \mnras, 310, 1087

\bibitem[\protect\citeauthoryear{{Spergel}, {Bean}, {Dor{\'e}}, {Nolta},
  {Bennett}, {Dunkley}, {Hinshaw}, {Jarosik}, {Komatsu}, {Page}, {Peiris} \&
  {Verde}}{{Spergel} et~al.}{2007}]{Spergel2007}
{Spergel} D.~N.,  {Bean} R.,  {Dor{\'e}} O.,  {Nolta} M.~R.,  {Bennett} C.~L.,
  {Dunkley} J.,  {Hinshaw} G.,  {Jarosik} N.,  {Komatsu} E.,  {Page} L.,
  {Peiris} H.~V.,    {Verde} L.,  2007, \apjs, 170, 377

\bibitem[\protect\citeauthoryear{{Stinson}, {Seth}, {Katz}, {Wadsley},
  {Governato} \& {Quinn}}{{Stinson} et~al.}{2006}]{Stinson2006}
{Stinson} G.,  {Seth} A.,  {Katz} N.,  {Wadsley} J.,  {Governato} F.,
  {Quinn} T.,  2006, \mnras, 373, 1074

\bibitem[\protect\citeauthoryear{{Stinson}, {Bailin}, {Couchman}, {Wadsley},
  {Shen}, {Nickerson}, {Brook} \& {Quinn}}{{Stinson}
  et~al.}{2010}]{Stinson2010}
{Stinson} G.~S.,  {Bailin} J.,  {Couchman} H.,  {Wadsley} J.,  {Shen} S.,
  {Nickerson} S.,  {Brook} C.,    {Quinn} T.,  2010, \mnras, 408, 812

\bibitem[\protect\citeauthoryear{{Stinson}, {Brook}, {Macci{\`o}}, {Wadsley},
  {Quinn} \& {Couchman}}{{Stinson} et~al.}{2013}]{Stinson2013}
{Stinson} G.~S.,  {Brook} C.,  {Macci{\`o}} A.~V.,  {Wadsley} J.,  {Quinn}
  T.~R.,    {Couchman} H.~M.~P.,  2013, \mnras, 428, 129

\bibitem[\protect\citeauthoryear{{Tacconi}, {Neri}, {Genzel}, {Combes},
  {Bolatto}, {Cooper}, {Wuyts}, {Bournaud}, {Burkert}, {Comerford}, {Cox},
  {Davis}, {F{\"o}rster Schreiber}, {Garc{\'{\i}}a-Burillo}, {Gracia-Carpio} \&
  {Lutz}}{{Tacconi} et~al.}{2013}]{Tacconi2013}
{Tacconi} L.~J.,  {Neri} R.,  {Genzel} R.,  {Combes} F.,  {Bolatto} A.,
  {Cooper} M.~C.,  {Wuyts} S.,  {Bournaud} F.,  {Burkert} A.,  {Comerford} J.,
  {Cox} P.,  {Davis} M.,  {F{\"o}rster Schreiber} N.~M.,
  {Garc{\'{\i}}a-Burillo} S.,  {Gracia-Carpio} J.,    {Lutz} D.,  2013, \apj,
  768, 74

\bibitem[\protect\citeauthoryear{{Tumlinson}, {Thom}, {Werk}, {Prochaska},
  {Tripp}, {Weinberg}, {Peeples}, {O'Meara}, {Oppenheimer}, {Meiring}, {Katz},
  {Dav{\'e}}, {Ford} \& {Sembach}}{{Tumlinson} et~al.}{2011}]{Tumlinson2011}
{Tumlinson} J.,  {Thom} C.,  {Werk} J.~K.,  {Prochaska} J.~X.,  {Tripp} T.~M.,
  {Weinberg} D.~H.,  {Peeples} M.~S.,  {O'Meara} J.~M.,  {Oppenheimer} B.~D.,
  {Meiring} J.~D.,  {Katz} N.~S.,  {Dav{\'e}} R.,  {Ford} A.~B.,    {Sembach}
  K.~R.,  2011, Science, 334, 948

\bibitem[\protect\citeauthoryear{{{\"U}bler}, {Naab}, {Oser}, {Aumer}, {Sales}
  \& {White}}{{{\"U}bler} et~al.}{2014}]{Ubler2014}
{{\"U}bler} H.,  {Naab} T.,  {Oser} L.,  {Aumer} M.,  {Sales} L.~V.,    {White}
  S.~D.~M.,  2014, \mnras, 443, 2092

\bibitem[\protect\citeauthoryear{{Vogelsberger}, {Genel}, {Springel}, {Torrey},
  {Sijacki}, {Xu}, {Snyder}, {Nelson} \& {Hernquist}}{{Vogelsberger}
  et~al.}{2014}]{Vogelsberger2014}
{Vogelsberger} M.,  {Genel} S.,  {Springel} V.,  {Torrey} P.,  {Sijacki} D.,
  {Xu} D.,  {Snyder} G.,  {Nelson} D.,    {Hernquist} L.,  2014, \mnras, 444,
  1518

\bibitem[\protect\citeauthoryear{{Wadsley}, {Stadel} \& {Quinn}}{{Wadsley}
  et~al.}{2004}]{Wadsley2004}
{Wadsley} J.~W.,  {Stadel} J.,    {Quinn} T.,  2004, New Astronomy, 9, 137

\bibitem[\protect\citeauthoryear{{Wadsley}, {Veeravalli} \&
  {Couchman}}{{Wadsley} et~al.}{2008}]{Wadsley2008}
{Wadsley} J.~W.,  {Veeravalli} G.,    {Couchman} H.~M.~P.,  2008, \mnras, 387,
  427

\bibitem[\protect\citeauthoryear{{Wang}, {Fu}, {Aumer}, {Kauffmann},
  {J{\'o}zsa}, {Serra}, {Huang}, {Brinchmann}, {van der Hulst} \&
  {Bigiel}}{{Wang} et~al.}{2014}]{Wang2014}
{Wang} J.,  {Fu} J.,  {Aumer} M.,  {Kauffmann} G.,  {J{\'o}zsa} G.~I.~G.,
  {Serra} P.,  {Huang} M.-l.,  {Brinchmann} J.,  {van der Hulst} T.,
  {Bigiel} F.,  2014, \mnras, 441, 2159

\bibitem[\protect\citeauthoryear{{Wang}, {Dutton}, {Stinson}, {Macci{\`o}},
  {Penzo}, {Kang}, {Keller} \& {Wadsley}}{{Wang} et~al.}{2015}]{Wang2015}
{Wang} L.,  {Dutton} A.~A.,  {Stinson} G.~S.,  {Macci{\`o}} A.~V.,  {Penzo} C.,
   {Kang} X.,  {Keller} B.~W.,    {Wadsley} J.,  2015, ArXiv e-prints

\bibitem[\protect\citeauthoryear{{Werk}, {Prochaska}, {Tumlinson}, {Peeples},
  {Tripp}, {Fox}, {Lehner}, {Thom}, {O'Meara}, {Ford}, {Bordoloi}, {Katz},
  {Tejos}, {Oppenheimer}, {Dav{\'e}} \& {Weinberg}}{{Werk}
  et~al.}{2014}]{Werk2014}
{Werk} J.~K.,  {Prochaska} J.~X.,  {Tumlinson} J.,  {Peeples} M.~S.,  {Tripp}
  T.~M.,  {Fox} A.~J.,  {Lehner} N.,  {Thom} C.,  {O'Meara} J.~M.,  {Ford}
  A.~B.,  {Bordoloi} R.,  {Katz} N.,  {Tejos} N.,  {Oppenheimer} B.~D.,
  {Dav{\'e}} R.,    {Weinberg} D.~H.,  2014, \apj, 792, 8

\bibitem[\protect\citeauthoryear{{White} \& {Frenk}}{{White} \&
  {Frenk}}{1991}]{White1991}
{White} S.~D.~M.,  {Frenk} C.~S.,  1991, \apj, 379, 52

\bibitem[\protect\citeauthoryear{{White} \& {Rees}}{{White} \&
  {Rees}}{1978}]{White1978}
{White} S.~D.~M.,  {Rees} M.~J.,  1978, \mnras, 183, 341

\end{thebibliography}

\clearpage

\end{document}